\documentclass[compsoc,conference,a4paper,10pt,times]{IEEEtran}
\usepackage{cite}
\usepackage{textcomp}
\usepackage{bmpsize}
\usepackage{lipsum}
\usepackage[colorlinks=true,urlcolor=black]{hyperref}
\def\BibTeX{{\rm B\kern-.05em{\sc i\kern-.025em b}\kern-.08em
		T\kern-.1667em\lower.7ex\hbox{E}\kern-.125emX}}

\ifCLASSINFOpdf
\else
\fi

\usepackage{bm}
\usepackage{graphicx}
\usepackage{xcolor}
\usepackage{import}

\usepackage{paralist}

\usepackage{listings}
\usepackage[noend]{algpseudocode}
\usepackage{url}
\usepackage{multirow}
\usepackage[font={small,it}]{caption}
\usepackage{array}
\usepackage{rotating}   
\usepackage{makecell}
\usepackage{colortbl}
\usepackage{enumerate}
\usepackage{booktabs}
\usepackage{subfig}
\usepackage[ruled,vlined]{algorithm2e}
\usepackage{amssymb}
\usepackage{commath}
\usepackage{amsthm}
\usepackage{bm}
\usepackage{framed}

\usepackage[binary-units=true]{siunitx}
\usepackage{adjustbox}

\usepackage{balance}

\usepackage{titlesec}

\titlespacing*{\section}
{0pt}{0.3cm}{0.1cm}
\titlespacing*{\subsection}
{0pt}{0.3cm}{0.1cm}
\titlespacing*{\subsubsection}
{0pt}{0.2cm}{0.1cm}

\algrenewcommand\algorithmiccomment[1]{\hfill \textcolor{gray}{$\triangleright$ \textit{#1}}}

\newtheorem{theorem}{Theorem}

\newenvironment{proof1} {\vspace{-0.3cm}\begin{proof}[Proof (Informal)]} {\end{proof}}

\renewcommand{\paragraph}[1]{\vspace{0.1cm}\noindent{\bf #1.}}

\newcommand{\name}{\textsc{Aqua\-reum}\xspace}

\usepackage{amsmath}
\usepackage{amsfonts}

\usepackage{multicol}
\usepackage{comment}

\usepackage{lipsum}

\usepackage{mdframed}
\usepackage{xifthen}
\usepackage{diagbox}
\usepackage{longtable}
\usepackage{breakcites}

\usepackage{makecell}
\newcolumntype{x}[1]{>{\centering\arraybackslash}p{#1}}

\usepackage{ragged2e}

\usepackage{tikz}

\usepackage{etoolbox}  \newcommand{\repthanks}[1]{\textsuperscript{\ref{#1}}}
\makeatletter
\patchcmd{\maketitle}
  {\def\thanks}
  {\let\repthanks\repthanksunskip\def\thanks}
  {}{}
\patchcmd{\@maketitle}
  {\def\thanks}
  {\let\repthanks\@gobble\def\thanks}
  {}{}
\newcommand\repthanksunskip[1]{\unskip{}}
\makeatother

\newcommand{\N}{{{\mathbb N}}}

\def\bits{\{0,1\}}

\def\union{\cup}

\newcommand{\secpar}{\kappa}

\newcommand{\suc}{\ensuremath{\mathcal{S}}\xspace}
\newcommand{\env}{\ensuremath{\mathcal{Z}}\xspace}
\newcommand{\adv}{\ensuremath{\mathcal{A}}\xspace}

\newcommand{\msg}[2]{\ifthenelse{\isempty{#1}}{\ensuremath{(sid, #2)}}{\ifthenelse{\isempty{#2}}{\ensuremath{(\textsc{#1}, sid)}}{\ensuremath{(\textsc{#1}, sid, #2)}}}}

\newcommand{\Fuc}{\ensuremath{\mathcal{F}\xspace}}

\newcommand{\IAdv}{\suc}

\newcommand{\rel}{\ensuremath{\mathcal{R}}\xspace}

\newcommand{\nizk}{\ensuremath{\mathsf{NIZK}}\xspace}

\newcommand{\nizkrel}[2]{\ifthenelse{\isempty{#1}}{\ensuremath{\nizk_{\rel}.{#2}}\xspace}{\ensuremath{\nizk_{#1}.{#2}}\xspace}}

\newcommand{\ie}{i.e.,\xspace}
\newcommand{\eg}{e.g.,\xspace}

\newcommand{\AQledger}{\Fuc_{\ledger}}

\newcommand{\ledger}{L}

\newcommand{\submQrySet}{\mathsf{Q_S}}
\newcommand{\censQrySet}{\mathsf{Q_C}}
\newcommand{\submTxSet}{\mathsf{T_S}}
\newcommand{\submRecSet}{\mathsf{R_S}}
\newcommand{\submRecNotHndSet}{\mathsf{R_{NH}}}
\newcommand{\submQryNotHndSet}{\mathsf{Q_{NH}}}
\newcommand{\ledgerStateList}{L}
\newcommand{\ledgerStateListPub}{\ledgerStateList_{pb}}
\newcommand{\ledgerState}{\mathcal{L}}
\newcommand{\censoredTx}{\mathsf{T_C}}
\newcommand{\opCorrupt}{\mathsf{Corr}}
\newcommand{\tx}{\mathsf{tx}}

\newcommand{\proc}{\mathtt{proc}}
\newcommand{\handle}{\mathtt{proc}}

\newcommand{\action}{\mathsf{action}}
\newcommand{\batch}{\mathsf{batch}}

\newcommand{\client}{\mathbb{C}}
\newcommand{\op}{\mathbb{O}}
\newcommand{\qry}{\mathsf{qry}}

\newcommand{\proto}{\Pi}
\newcommand{\zenv}{\mathcal{Z}}

\newcommand{\rand}{r}
\newcommand{\envinput}{z}

\newcommand{\func}{\mathcal{F}}

\newcommand{\funcmodel}{\mathcal{F}}

\newcommand{\realmodel}{{\mathsf{REAL}_{\proto, \adv, \zenv}}}
\newcommand{\ensreal}{{\{\mathsf{REAL}_{{\proto}, \adv, \zenv}(\secpar,\envinput,\rand)\}_{\secpar \in \N,\envinput \in \bits^{*}}}}
\newcommand{\realvar}{{\mathsf{REAL}_{{\proto}, \adv, \zenv}}}

\newcommand{\idealmodel}{{\mathsf{IDEAL}_{\funcmodel, \IAdv, \zenv}}}

\newcommand{\ensideal}{{\{\mathsf{IDEAL}_{\funcmodel, \IAdv, \zenv}(\secpar,\envinput,\rand)\}_{\secpar \in \N, \envinput \in \bits^{*}}}}

\newcommand{\idealvar}{{\mathsf{IDEAL}_{\funcmodel, \IAdv, \zenv}}}

\newcommand{\myhalfbox}[5]{
    \begin{figure}[tpb]
        \vspace{-0.2cm}
        \centering
    \begin{tikzpicture}
        \node[anchor=text,text width=\columnwidth-0.6cm, draw, rounded corners, line width=1pt, fill=#3, inner sep=2.5mm] (big) {\\#4};
        \node[draw, rounded corners, line width=.5pt, fill=#2, anchor=west, xshift=5mm] (small) at (big.north west) {#1};
    \end{tikzpicture}
	\vspace{-0.4cm}
    \caption{#5}
    \vspace{-0.4cm}
    \end{figure}
}

\begin{document}

\date{}

\title{\Large \bf \name: Non-Equivocating Censorship-Evident Centralized Ledger with EVM-Based Verifiable Execution using Trusted Computing and Blockchain}

	\author{\IEEEauthorblockN{Ivan Homoliak}
	\IEEEauthorblockA{
		\textit{Brno University of Technology}\\
		\textit{Faculty of Information Technology}\\
		homoliak@fit.vutbr.cz \vspace{-0.5cm}}
	\IEEEauthorblockN{\hspace{1.5cm}}
	\and				
	\IEEEauthorblockN{Mario Larangeira}
	\IEEEauthorblockA{
		\textit{Tokyo Institute of Technology / IOG}\\
		mario.larangeira@iohk.io \vspace{-0.5cm}}
	\IEEEauthorblockN{\hspace{1.5cm}}
	\and				
	\IEEEauthorblockN{Martin Pere\v{s}\'{i}ni}
	\IEEEauthorblockA{
		\textit{Brno University of Technology}\\
		\textit{Faculty of Information Technology}\\
		iperesini@fit.vutbr.cz \vspace{-0.5cm}}
	\IEEEauthorblockN{\hspace{1.5cm}}
	\and					
	\IEEEauthorblockN{Pawel Szalachowski}
	\IEEEauthorblockA{
		\textit{Singapore University of Technology and Design}\\
		pjszal@gmail.com}			
}

\IEEEoverridecommandlockouts

\maketitle

\begin{abstract}
Distributed ledger systems (i.e., blockchains) have received a lot of
attention. They promise to enable mutually untrusted participants
to execute transactions while providing the immutability of the data and censorship resistance.    Although decentralized ledgers are a disruptive innovation, as of today, they suffer from scalability,
privacy, or governance issues.  Therefore, they are inapplicable for many
important use cases, where interestingly, centralized ledger systems might
gain adoption. 
Unfortunately, centralized ledgers
have also drawbacks, e.g., a lack of efficient verifiability or a higher risk of censorship and equivocation.

In this paper, we present \name, a novel framework for centralized ledgers
removing their main limitations.  By a unique combination of a trusted execution environment (TEE) with a public blockchain, \name provides publicly
verifiable non-equivocating censorship-evident private and high-performance
ledgers. 
\name is integrated with a Turing-complete virtual machine (e.g., EVM),
allowing arbitrary transaction processing logic, such as transfers or client-specified smart contracts.  \name is fully implemented and can process over 400 transactions per second on a commodity PC.
Furthermore, we modeled \name using the Universal Composability framework and proved its security. 
\end{abstract}

\section{Introduction}
\label{sec:intro}

Ledger systems are append-only databases providing immutability (i.e., tamper
resistance) as a core property. 
To facilitate their append-only feature, cryptographic
constructions, such as hash chains~\cite{haber1990time}, Merkle trees~\cite{bayer1993improving} and history trees~\cite{crosby2009efficient}, are usually deployed.
Traditionally, public ledger systems are centralized and controlled by a single
entity that acts as a trusted party.  In such a setting, ledgers are deployed in various applications, including logging~\cite{amazon-qldb,ledgerdb,dowling2016secure,tomescu2019transparency}, event stores~\cite{aws-event-store}
timestamping services~\cite{ledgerdb},
payments~\cite{institutions2004wholesale,carter2023payment,folkerts1997wholesale}, repositories~\cite{github-commits}, or public logs of various artifacts (e.g.,
keys~\cite{melara2015coniks,chase2019seemless}, binaries~\cite{fahl2014hey}, and certificates issued~\cite{laurie2013certificate} or revoked~\cite{laurie2012revocation} by authorities).

Although being successfully deployed and envisioned for multiple novel use
cases, centralized ledgers have some fundamental limitations due to their
centralization.  Firstly, they lack efficient verifiability, which would ensure
their clients that the ledger is indeed append-only and internally consistent 
(i.e., without conflicting transactions).  A naive solution is just to
publish the ledger or share it with parties interested in auditing it, which, however,
may be inefficient or stand against the ledger operator's deployment models (e.g., the privacy of the clients conducting financial transactions 
can be violated).
Second, it is challenging to provide non-equivocation to centralized
systems~\cite{mazieres2002building,dowling2016secure}.  
In simple yet devastating fork attacks, a ledger operator creates two conflicting copies of the ledger and presents it to different clients.  
Although the forked ledgers are internally consistent, the ``global'' view of the database is equivocated, thus completely
undermining the security of the entire system.  Finally, centralized systems are
inherently prone to censorship.  A ledger operator can refuse any request or transaction at her will without leaving any evidence of censoring.  This may be risky especially when a censored client may suffer from some consequences (e.g.,
fines when being unable to settle a transaction on time) or in the case when
the operator wishes to hide some ledger content (e.g., data proving her misbehavior).
On the other hand, public distributed ledgers (e.g.,~\cite{nakamoto2008bitcoin,wood2014ethereum,gilad2017algorand,kiayias2017ouroboros}) combine an append-only
cryptographic data structure with a consensus algorithm, spreading trust across
all participating consensus nodes.  These systems are by design publicly
verifiable, non-equivocating, and censorship-resistant. 
However, they offer low throughput, are expensive, do not inherently provide privacy, and their public nature makes their governance difficult or unacceptable for many applications.

Another line of works focuses on non-equivocation by utilizing gossip protocols~\cite{chuat2015efficient,dahlberg2018aggregationbased}, consensus of auditing nodes~\cite{kim2013accountable,basin2014arpki,paccagnella2020custos}, and blockchains~\cite{tomescu2017catena,guarnizo2019pdfs}.
However, they do not guarantee correct execution, which motivates our work.
Some approaches combine blockchain with TEE to improve lacking properties of blockchain, such as privacy (e.g.,~\cite{secret-ppp,phala-ppp,oasis-ppp,obscuro-ppp,integritee-ppp}), throughput (e.g.,~\cite{lind2017teechain,cheng2018ekiden,das2019fastkitten,FrassettoJKKSFS23}), and adding Turing-complete smart contract functionality (e.g.,~\cite{das2019fastkitten}). 
Nevertheless, these approaches are in principle decentralized, and thus they impose on-chain expenses on clients.
In contrast, this work focuses on a centralized design that should provide almost the same features as decentralized ones but with additional privacy and at much lower costs clients pay.

\paragraph{\textbf{Proposed Approach}}
In this paper, we propose \name, a framework for centralized ledgers mitigating
their main limitations.  \name introduces a unique combination of TEE and a public smart contract platform to provide verifiability, non-equivocation, and censorship mitigation.  
In \name, a ledger operator deploys a pre-defined TEE enclave code, which verifies the consistency and correctness of the ledger for every ledger update. 
Then, a proof produced by
the enclave is published through a public smart contract platform, guaranteeing that no alternative snapshot of this ledger exists.  
Furthermore, whenever a client suspects that her request (or transaction) is censored, she can (confidentially) request a resolution via the smart contract platform.

\name can be adjusted to different ledgers and use cases, but we implemented and deployed it with customized enclave-ready Ethereum Virtual Machine (EVM) from Microsoft Confidentiality Framework~\cite{eEVM-Microsoft} since EVM provides a Turing-complete execution environment and it is widely adopted in many permissionless~\cite{evm-blockchains-2021} and permissioned~\cite{nevile2022enterprise} blockchains. 
\name enables hosting of arbitrary ledger applications, such as key-value stores, tokens, or client-defined smart contracts.\footnote{Note that we demonstrate its application for Central Bank Digital Currency (CBDC) in Submission \#71.}  
\section{Background}
\label{sec:pre}
\subsection{Blockchain and Smart Contracts}
A blockchain (a.k.a., a distributed ledger) is an append-only data structure
that is resistant by design against modifications combined with a consensus
protocol.
In a blockchain, blocks containing data records are linked using a cryptographic hash function, and each new block has to be agreed upon by participants running a consensus protocol (i.e., \textit{consensus nodes}).
Each block may contain data records representing orders that transfer crypto-tokens, application codes written in a platform-supported language, and the execution orders of such application codes. 
These application codes are referred to as \textit{smart contracts}, and they encode arbitrary processing logic (e.g., agreements) written in the supported language of a smart contract platform.
Interactions between clients and the smart contract platform are based on messages called \textit{transactions}, which can contain either orders transferring crypto-tokens or calls of smart contract functions.
All transactions sent to a blockchain are validated by consensus nodes who maintain a replicated state of the blockchain.
To incentivize consensus nodes, blockchain platforms introduce reward and fee schemes.

\subsection{Trusted Execution Environment}
Trusted Execution Environment (TEE) is a hardware-based component that can securely execute arbitrary code in an isolated environment. 
TEE uses cryptography primitives and hardware-embedded secrets that protect data confidentiality and the integrity of computations.
In particular, the adversary model of TEE usually includes privileged applications and an operating system, which may compromise unprivileged user-space applications.
There are several practical instances of TEE, such as Intel Software Guard Extensions (SGX)~\cite{anati2013innovative,mckeen2013innovative,hoekstra2013using} available at Intel's CPUs or based on RISC-V architecture such as Keystone-enclave~\cite{Keystone-enclave} and Sanctum~\cite{costan2016sanctum}.
In the context of this work, we build on top of Intel SGX, therefore we adopt the terminology introduced by it.

Intel SGX is a set of instructions that ensures hardware-level isolation of protected user-space codes called \textit{enclaves}. 
An enclave process cannot execute system calls but can read and write memory outside the enclave. 
Thus isolated execution in SGX may be viewed as an ideal model in which a process is guaranteed to be executed correctly with ideal confidentiality, while it might run on a potentially malicious operating system.

Intel SGX allows a local process or a remote system to securely communicate with the enclave as well as execute verification of the integrity of the enclave's code. 
When an enclave is created, the CPU outputs a report of its initial state, also referred to as a \textit{measurement}, which is signed by the private key of TEE and encrypted by a public key of Intel Attestation Service (IAS). 
The hardware-protected signature serves as proof that the measured code is running in an SGX-protected enclave, while the encryption by IAS public key ensures that the SGX-equipped CPU is genuine and was manufactured by Intel.

\subsection{Merkle Tree}\label{sec:MT-background}
A Merkle tree~\cite{merkle1989certified} is a data structure based on the binary tree in which each leaf node contains a hash of a single data block, while each non-leaf node contains a hash of its concatenated children.
At the top of a Merkle tree is the root hash, which provides a tamper-evident summary of the contents.
A Merkle tree enables efficient verification as to whether some data are associated with a leaf node by comparing the expected root hash of a tree with the one computed from a hash of the data in the query and the remaining nodes required to reconstruct the root hash (i.e., \textit{proof} or \textit{authentication path}).
The reconstruction of the root hash has logarithmic time and space complexity, which makes the Merkle tree an efficient scheme for membership verification.
To provide a membership verification of element $x_i$ in the list of elements $X = \{x_i\}, i \geq 1$, the Merkle tree supports the following operations:
\begin{compactdesc}
	\item{$\mathbf{MkRoot(X) \rightarrow  Root}$:} an aggregation of all elements of the list $X$ by a Merkle tree, providing a single value $Root$. 
	
	\item{$\mathbf{MkProof(x_i, X) \rightarrow  \pi^{mk}}$:} a Merkle proof generation for the $i$th element $x_i$ present in the list of all elements $X$. 
	
	\item{$\mathbf{\pi^{mk}.Verify(x_i, Root) \rightarrow  \{True, False\}}$:}  verification of the Merkle proof $\pi^{mk}$, witnessing that $x_i$ is included in the list $X$ that is aggregated by the Merkle tree with the root hash  $Root$.
\end{compactdesc}

\subsection{History Tree}\label{sec:background-historyT} 
A Merkle tree has been primarily used for proving membership.
However, Crosby and Wallach~\cite{crosby2009efficient} extended its application for an append-only tamper-evident log, denoted as a \textit{history tree}.
A history tree is the Merkle tree, in which leaf nodes are added in an append-only fashion, allowing to production of logarithmic proofs witnessing that arbitrary two versions of the tree are consistent (i.e., one version of the tree is an extension of another).
Therefore, once added, a leaf node cannot be modified nor removed. 
A history tree brings a versioned computation of hashes over the Merkle tree, enabling it to prove that different versions (i.e., commitments) of a log, with distinct root hashes, make consistent claims about the past.
To provide a tamper-evident history system~\cite{crosby2009efficient}, the log represented by the history tree $L$ supports the following operations:
\begin{compactdesc}
	\item[$\mathbf{L.add(x) \rightarrow C_j}$:] appending of the record $x$ to $L$, returning a new commitment $C_j$ that represents the most recent value of the root hash of the history tree.
	
	\item[$\mathbf{L.IncProof(C_i, C_j) \rightarrow \pi^{inc}}$:] an incremental proof generation between 2 commitments, where $i \leq j$.
	
	\item[$\mathbf{L.MemProof(i, C_j) \rightarrow \pi^{mem} }$:] a membership proof generation for the record $x_i$ from the commitment $C_j$, where $i \leq j$. 
	
	\item[$\mathbf{\pi^{inc}.Verify(C_i, C_j) \rightarrow \{True, False\}}$:]  verification of the incremental proof $\pi^{inc}$, witnessing that $C_j$ contains the same history of records $x_k, k \in \{0,\ldots,i\}$ as $C_i$, where $i \leq j$.  
	
	\item[$\mathbf{\pi^{mem}.Verify(i, x_i, C_j) \rightarrow \{True, False\}}$:]  verification of the membership proof $\pi^{mem}$, witnessing that $x_i$ is the $i$th record in the $j$th version of $L$, fixed by the commitment $C_j$,  $i \leq j$. 
	
	\item[$\mathbf{\pi^{inc}.DeriveNewRoot() \rightarrow C_j}$:] a reconstruction of the commitment $C_j$ from the incremental proof $\pi^{inc}$ that was generated by $L.IncProof(C_i, C_j)$.
	
	\item[$\mathbf{\pi^{inc}.DeriveOldRoot() \rightarrow C_i}$:] a reconstruction of the commitment $C_i$ from the incremental proof $\pi^{inc}$ that was generated by $L.IncProof(C_i, C_j)$.
	
\end{compactdesc}

\subsection	{Notation}
The notation used throughout the paper is presented in the following.
By $\{msg\}_\mathbb{U}$, we denote the message $msg$ digitally signed by $\mathbb{U}$, and by $msg.\sigma$ we refer to a signature;
$h(.)$ stands for a cryptographic hash function;
$\|$ is the string concatenation;
$\%$ represents modulo operation over integers; 
$\Sigma_{p}. \{KeyGen, Verify, Sign\}$ represents a signature (and encryption) scheme of the platform $p$, where $p \in \{pb, tee\}$ (i.e., a public blockchain platform and TEE platform);
and $SK_\mathbb{U}^{p}$, $PK_\mathbb{U}^{p}$ is the private/public key-pair of $\mathbb{U}$, under $\Sigma_{p}$.
Then, we use $\pi^{s}$ for denoting proofs of various data structures $s \in \{mk, mem, inc\}$: 
$\pi^{mk}$ denotes the inclusion proof in the Merkle tree, 
$\pi^{mem}$ and $\pi^{inc}$ denote the membership proof and the incremental proof in the history tree, respectively.

\section{System Model and Overview}
\label{sec:overview}

\subsection{System Model}
In \name, an \textit{operator} is an entity that maintains and manages a ledger containing chronologically sorted transactions. 
\textit{Clients} interact with the ledger by sending requests, such as queries and transactions to be handled.  
We assume that all involved parties can interact with a blockchain platform supporting smart contracts (e.g., EVM-compatible blockchains). 
Next, we assume that the operator has access to a TEE platform (e.g., Intel SGX).
Finally, we assume an adversary $\adv$  who controls the operator and her goals are as follows:
\begin{compactdesc}
	\item \textbf{Violation of the ledger's integrity} by creating its
	internal inconsistent state -- e.g., via inserting 2 conflicting
	transactions or by removing/modifying existing transactions.
	\item \textbf{Equivocation of the ledger} by presenting at least two inconsistent views of the ledger to (at least) two distinct clients who would accept such views as valid.
	\item \textbf{Censorship of client requests} without leaving any audit trails evincing the censorship occurrence.
\end{compactdesc}
Next, we assume that $\adv$ cannot undermine the cryptographic primitives used, 
the underlying blockchain platform, and the TEE platform deployed.

\subsection{Desired Properties}\label{sec:desired-properties}
We target the following security properties:
\begin{compactitem}
	\item[\textit{\textbf{Verifiability:}}] 
	clients should be able to obtain easily verifiable evidence that the
	ledger they interact with is internally \textit{correct} and \textit{consistent}. 
	In particular, it means that none of the previously inserted transactions was neither modified nor deleted, and there are no conflicting transactions.  
Besides, the system should be \textit{self-auditable}, such that any client can verify that some transaction is included in the ledger. 

	\item[\textit{\textbf{Non-Equivocation:}}]
	the system should protect from forking attacks and thus guarantee that no
	concurrent versions of the ledger exist at any point in
	time.

	\item[\textit{\textbf{Censorship Evidence:}}] 
	detecting censorship in a centralized system is 
	challenging, as its operator can simply pretend unavailability to censor undesired queries or transactions.  
We emphasize that proving censorship is
	a non-trivial task since it is difficult to distinguish ``pretended''
	unavailability from ``genuine'' one. 
\end{compactitem}
Next, we intend our system to provide \textit{privacy}
(keeping the clients' communication confidential), \textit{efficiency}, \textit{deployability} with today's technologies and infrastructures, and
\textit{flexibility} enabling various use cases.

\begin{figure}[t]
	\centering
	\includegraphics[width=0.44\textwidth]{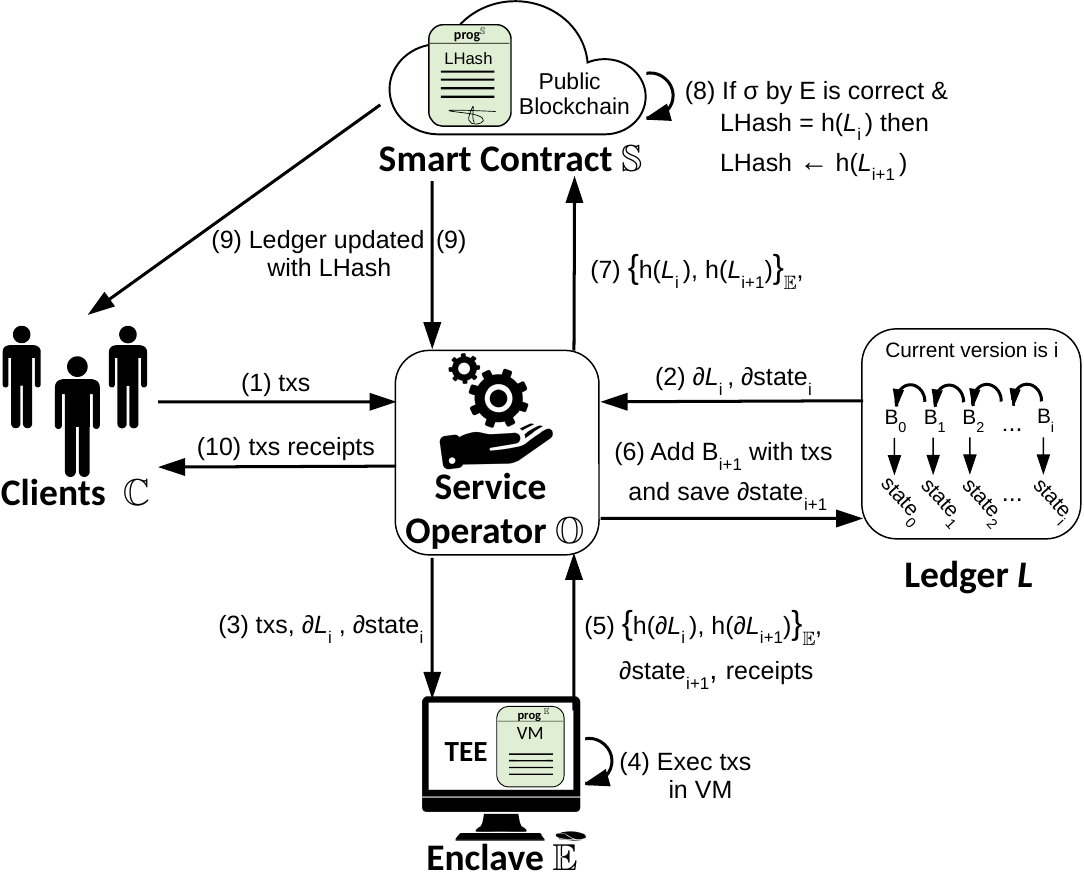} 
	\vspace{-0.1cm}
	\caption{Operation procedure of \name ledger.}
	\label{fig:overview-normal-op}
	\vspace{-0.4cm}
\end{figure}

\subsection{High-Level Overview}\label{sec:high-overview}
\subsubsection{\textbf{Setup}}
\name ledger is initialized by an operator ($\mathbb{O}$) who creates an internal ledger ($L$) that will store all transactions processed and the state that they render.  
Initially, $L$ contains an empty transaction set and a null state.
During the setup, $\mathbb{O}$ creates a TEE enclave ($\mathbb{E}$) whose role is to execute updates of $L$ and verify the consistency of $L$. Initialization of $\mathbb{E}$ involves the generation of two public private key pairs -- one for the signature scheme of TEE (i.e.,  $PK_{\mathbb{E}}^{tee}, SK_{\mathbb{E}}^{tee}$) and one for the signature scheme of the public blockchain (i.e., $PK_{\mathbb{E}}^{pb}, SK_{\mathbb{E}}^{pb}$).\footnote{Note that neither of the private keys ever leaves $\mathbb{E}$.}
The code of $\mathbb{E}$ is public (see \autoref{alg:enclave-VM} and \autoref{alg:enclave-VM-cens}), and it can be remotely attested with the TEE infrastructure by any client ($\mathbb{C}$).
Next, $\mathbb{O}$ generates her public-private key pair (i.e., $PK_{\mathbb{O}}, SK_{\mathbb{O}}$) and deploys a special smart contract ($\mathbb{S}$) initialized with the empty $L$ represented by its hash $LHash$, the $\mathbb{O}$'s public key $PK_{\mathbb{O}}$, and both enclave public keys $PK_{\mathbb{E}}^{tee}$ and $PK_{\mathbb{E}}^{pb}$.
After the deployment of $\mathbb{S}$, an instance of $L$ is uniquely identified by the address of~$\mathbb{S}$.
A ($\mathbb{C}$) wishing to interact with $L$ obtains the address of $\mathbb{S}$ and performs the remote attestation of $\mathbb{E}$ using the $PK_{\mathbb{E}}^{tee}$.

\subsubsection{\textbf{Operation}}
Whenever $\mathbb{C}$ sends a transaction $tx$ to $\mathbb{O}$, she relays it to $\mathbb{E}$, which validates whether $tx$ is authentic and non-conflicting; and if so, $\mathbb{E}$ executes $tx$ in its virtual machine (VM) and updates $L$ with it, resulting into the new version of $L$. 
Next, $\mathbb{O}$ responds to $\mathbb{C}$ with an execution \textit{receipt} of $tx$ and \textit{a version transition} of $L$, both signed by $\mathbb{E}$, proving that $tx$ was processed successfully and is included in the new version. 

\paragraph{Ledger Update Procedure}
For efficiency reasons, transactions are processed in \textit{batches} (i.e., \textit{blocks}).  
In detail, $\mathbb{O}$ updates $L$ from version $i$ to $i+1$ as follows (see \autoref{fig:overview-normal-op}):

\begin{compactenum}[a)]
	\item $\mathbb{O}$ relays all received transactions since the
	previous update to $\mathbb{E}$, together with the current partial state of $L$ denoted as $\partial state_i$, such that $h(\partial state_i) = h(state_i)$, and partial ledger $\partial L_i$, such that $h(\partial L_i) = h(L_i)$, which are required to validate $L$'s consistency and do its incremental extension.
	
	\item $\mathbb{E}$ validates and executes the transactions in its VM, updates $\partial state_i$ and $\partial L_i$ to their new versions $\partial state_{i+1}$ and $\partial L_{i+1}$, respectively. 
	Finally $\mathbb{E}$ creates a blockchain transaction $\{h(\partial L_i), h(\partial L_{i+1})\}_\mathbb{E}$ signed by $SK_{\mathbb{E}}^{pb}$,\footnote{Note that $\{h(\partial L_i), h(\partial L_{i+1})\}_\mathbb{E} = \{h(L_i), h(L_{i+1})\}_\mathbb{E}$ } which represents a version transition of $L$ from version $i$ to version $i+1$, also referred to as the \textbf{version transition pair}.
	
	\item The blockchain transaction with version transition pair is returned to $\mathbb{O}$ who sends this transaction to $\mathbb{S}$. 
	
	\item $\mathbb{S}$ accepts the second item of the pair as the current hash of $L$ iff it is signed by $SK_{\mathbb{E}}^{pb}$ and the current hash of $L$ stored by $\mathbb{S}$ (i.e., $LHash$) equals to the first item of the pair.  
\end{compactenum}

\smallskip\noindent
After the update of $L$ is finished, $\mathbb{C}$s with receipts obtained can verify that their transactions were processed by $\mathbb{E}$ (see details in \autoref{sec:receipt-retrieval}).  
The update procedure ensures that the new version of $L$ is:
(1) \textbf{internally correct} since it was executed by trusted code of $\mathbb{E}$, 
(2) a \textbf{consistent} extension of the previous version -- relying on trusted code of $\mathbb{E}$ and a witnessed version transition by $\mathbb{S}$, and 
(3) \textbf{non-equivocating} since $\mathbb{S}$ stores only hash of a single version of $L$ (i.e., $LHash$). 

Whenever $\mathbb{C}$ detects that its transactions or read queries are censored, $\mathbb{C}$ might send such requests via $\mathbb{S}$ (see details in \autoref{sec:censored-transaction} and \autoref{sec:censored-query}). 
To do so, $\mathbb{C}$ encrypts her request with $PK_{\mathbb{E}}^{pb}$ and publishes it on the blockchain via $\mathbb{S}$. 
$\mathbb{O}$ noticing a new request is obligated to pass the request to $\mathbb{E}$, which will process the request and reply (via $\mathbb{S}$) with a response encrypted by $PK_{\mathbb{C}}^{pb}$.  
If a pending request at $\mathbb{S}$ is not handled by $\mathbb{O}$, it is public evidence that $\mathbb{O}$ censors it. 
We do not specify how $\mathbb{C}$  utilize such evidence, but we envision it to be shown in a legal dispute or $\mathbb{S}$ could have automated deposit-based punishment rules.
Note that censorship resolution can be transparent to the user at $\mathbb{C}$. 

\begin{figure}[t!]
	\centering
\includegraphics[width=0.49\textwidth]{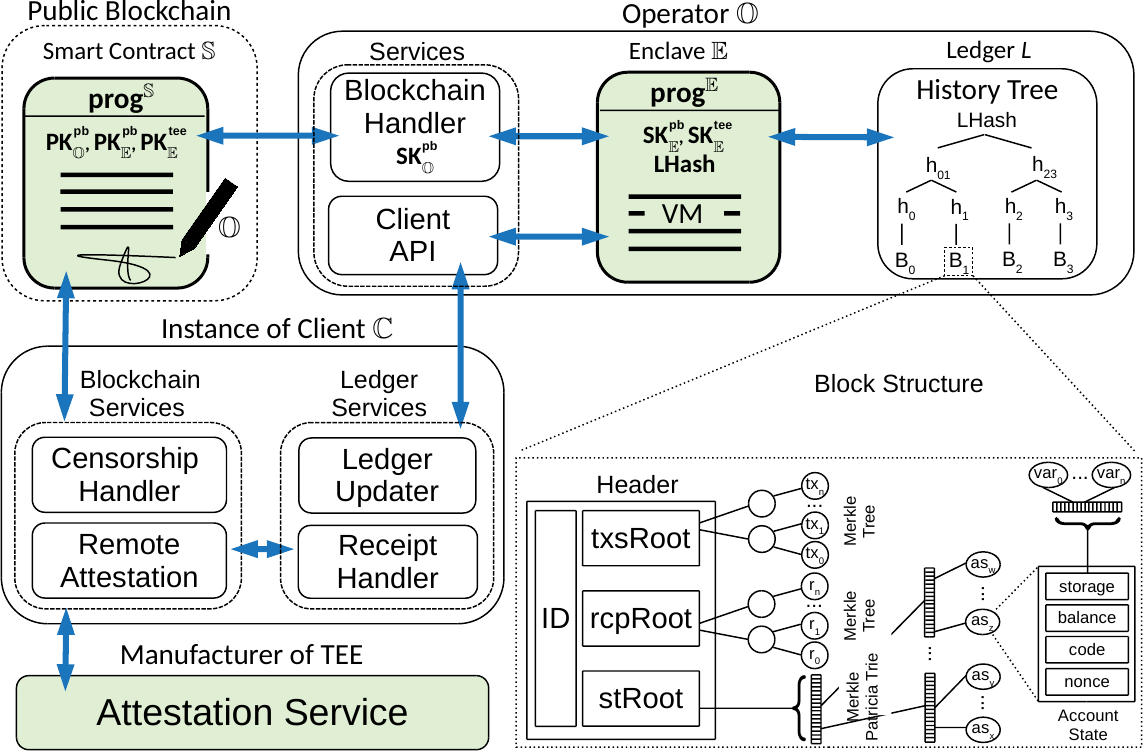} 
	\vspace{-0.3cm}
	\caption{\name components. Trusted ones are in green.}
	\label{fig:overview-details}		
	\vspace{-0.4cm}
\end{figure}

\subsection{Design Consideration}\label{sec:design-considerations}
We might design $L$ as an append-only chain (as in blockchains), but such a design would bring a high overhead on clients who want to verify that a particular block belongs to $L$.
During the verification, clients would have to download the headers of all blocks between the head of $L$ and the block in the query, resulting in linear space \& time complexity.
In contrast, when a history tree (see \autoref{sec:background-historyT}) is utilized for integrity preservation of $L$, the presence of any block in $L$ can be verified with logarithmic space and time complexity. 
\section{Details of \name}
\label{sec:details}

The schematic overview of \name is depicted in \autoref{fig:overview-details}. The right part of the figure describes data aggregation of $L$.
We utilize a history tree~\cite{crosby2009efficient} for tamper-evident logging of data blocks due to its efficient membership and incremental proofs (see \autoref{sec:design-considerations}).
The aggregation of blocks within a history tree is represented by root hash $LRoot$, which instantiates ledger hash $LHash$ from \autoref{sec:high-overview}. 
In \name, each data block consists of a header, a list of transactions, and a list of execution receipts from the VM that is running within $\mathbb{E}$.
A header contains the following fields:
\begin{compactitem}
	
	\item \textbf{ID}: this field is assigned for each newly created block as a counter of all blocks.
	ID of each block represents the $ID$th version of the history tree of $L$, which contains blocks $B_0,\ldots, B_{ID-1}$ and is characterized by the root hash $r \leftarrow MkRoot(\{H_0,\ldots, H_{ID-1}\})$, where $H_i$ stands for a header of a block $B_i$.	
	Note that the $ID$th version of $L$ with the root hash $r$ can also be expressed by the notation $\#(r)$.
	
	\item \textbf{txsRoot}, \textbf{rcpRoot}: root hash values that aggregate a set of transactions and the set of their corresponding execution receipts (containing execution logs) by Merkle trees~\cite{merkle1989certified},
	
	\item \textbf{stRoot}: the root hash that aggregates the current global state of the virtual machine by Merkle-Patricia Trie (MPT)~\cite{wood2014ethereum,Merkle-Patricia-Trie-eth}.\footnote{See optional background on MPT in Appendix~\ref{appendix:mpt-background}.} 
	In detail, MPT aggregates all account states into a global state, where keys of MPT represent IDs of client accounts (i.e., $h(PK_{\mathbb{C}}^{pb})$) and values represent an account state data structures, which (similar to~\cite{wood2014ethereum}) contains: 
	(1) \textit{balance} of a native token (if any),
	(2) \textit{code} that is executed when an account receives a transaction;
	accounts with no code represent simple accounts and accounts with a code field represent smart contract accounts,
	(3) \textit{nonce} represents the number of transactions sent from the simple account or the number of contracts created by the smart contract account,
	(4) \textit{storage} represents encoded variables of a smart contract, which can be realized by MPT~\cite{wood2014ethereum,Merkle-Patricia-Trie-eth} or other integrity-preserving mapping structures.
	
\end{compactitem}

\noindent
Although $\mathbb{O}$ persists the full content of $L$, she is unable to directly modify $L$ while remaining undetected since all modifications of $L$ must be made through $\mathbb{E}$.
In detail, upon receiving enough transactions from clients, $\mathbb{E}$ executes received transactions by its virtual machine (VM) and updates $L$. 
While updating $L$, $\mathbb{E}$ leverages the incremental proofs of the history tree to ensure integrity and consistency with the past versions of $L$.

$\mathbb{E}$ in our approach stores the last produced header ($hdr_{last}$) and the current root hash of the history tree of $L$ (i.e., $LRoot$), which enables $\mathbb{E}$ to make extensions of $L$ that are consistent with $L$'s history and at the same time avoiding dishonest $\mathbb{O}$ to tamper with $L$.
Although state-fullness of $\mathbb{E}$ might be seen as a limitation in the case of a failed enclave, we show how to deal with this situation and provide a procedure that publicly replaces a failed enclave using $\mathbb{S}$ (see \autoref{sec:failed-enclave}).

\begin{figure}[t]
\begin{center}
		\includegraphics[width=0.45\textwidth]{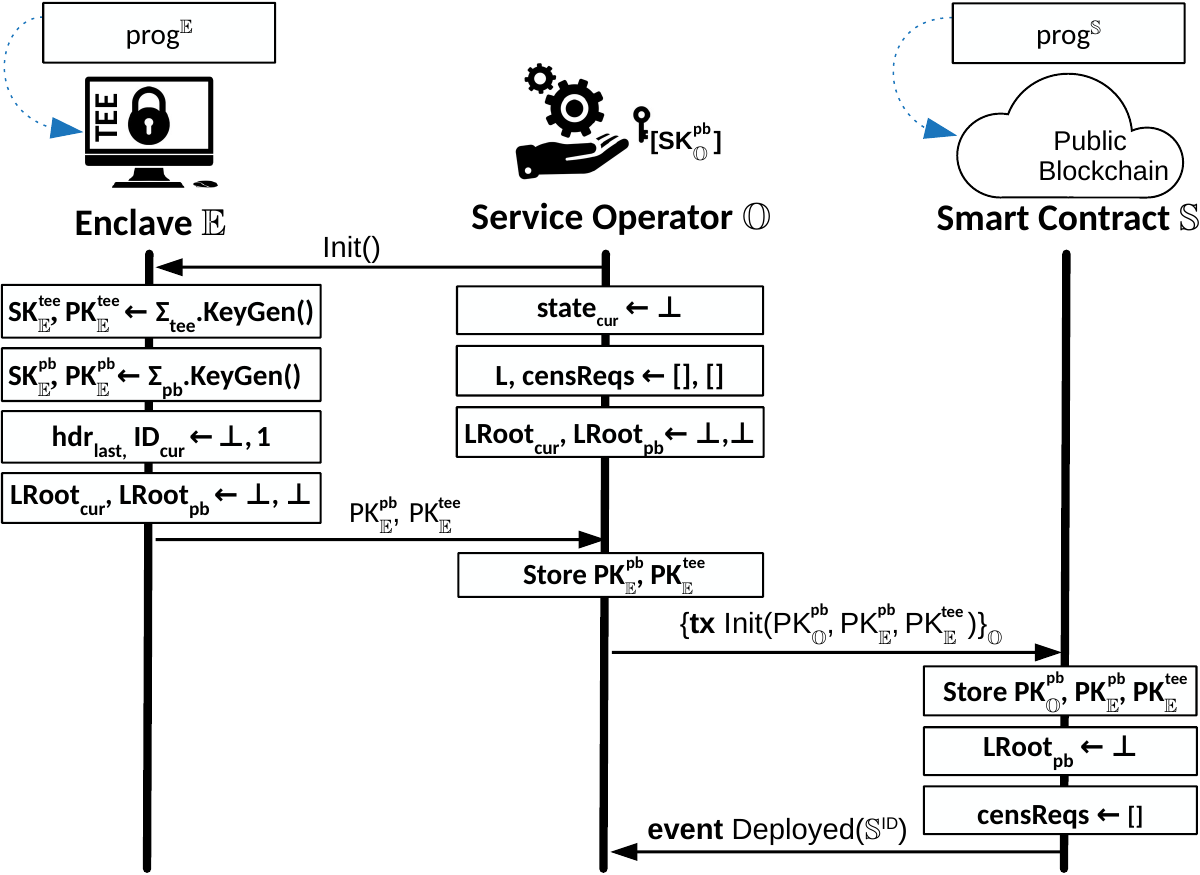} 
		\vspace{-0.1cm}
		\caption{Protocol for setup ($\Pi_{S}$).}		
		\label{fig:setup}
		\vspace{-0.6cm}
	\end{center}	
\end{figure}

\subsection{Setup}\label{sec:setup}
The setup of \name is presented in \autoref{fig:setup}.\footnote{We assume that $\mathbb{O}$ has already generated her key-pair $PK_\mathbb{O}^{pb}, SK_\mathbb{O}^{pb}$.} 
First, $\mathbb{O}$ initializes an empty $L$, a root hash $LRoot_{cur}$ for the most recent local version of $L$, the root hash $LRoot_{pb}$ for the version synchronized with PB, the empty global state of $L$, and the empty list of reported censored requests.
Then, $\mathbb{O}$ initializes $\mathbb{E}$ with code $prog^\mathbb{E}$ (see \autoref{alg:enclave-VM}). 
In this initialization, $\mathbb{E}$ generates two key-pairs, $SK_\mathbb{E}^{pb}, PK_\mathbb{E}^{pb}$ and $SK_\mathbb{E}^{tee}, PK_\mathbb{E}^{tee}$, respectively; 
the first key pair is intended for interaction with the blockchain platform and the second one is intended for the remote attestation with TEE infrastructure.
Next, $\mathbb{E}$ initializes $L$ and two root hashes in the same vein as $\mathbb{O}$ did. 
In addition, $\mathbb{E}$ stores the header $hdr_{cur}$ of the last block created and signed by $\mathbb{E}$ and its ID.
Then, $\mathbb{E}$ sends its public keys $PK_\mathbb{E}^{pb}$ and $PK_\mathbb{E}^{tee}$  to $\mathbb{O}$. 
Next, $\mathbb{O}$ creates a deployment transaction of $\mathbb{S}$'s code $prog^\mathbb{S}$ (see \autoref{alg:log-smart-contract}) with public keys $PK_\mathbb{E}^{pb}, PK_\mathbb{E}^{tee}$, $PK_\mathbb{O}$ as the arguments (see \autoref{alg:operator} of Appendix for pseudo-code of $\mathbb{O}$).
Then, $\mathbb{O}$ sends the deployment transaction to the blockchain.
In the constructor of $\mathbb{S}$, all public keys are stored, and the root hash of $L$ with the list of censored requests are initialized.
Finally, $\mathbb{S}$ publishes its identifier $\mathbb{S}^{ID}$, which serves as a public reference to~$\mathbb{S}$.

When the \name ledger is initialized, $\mathbb{C}$s register at $\mathbb{O}$. 
For simplicity, we omit details of the registration and access control, and we leave this up to the discretion of~$\mathbb{O}$.

\begin{algorithm}[t] 
	\scriptsize
	\SetKwProg{func}{function}{}{}
	
	$\triangleright$ \textsc{Declaration of types and functions:}\\
	\hspace{1em} \textbf{Header} \{$ID$, $txsRoot$, $rcpRoot$, $stRoot$\}; \\
	
	\hspace{1em} $\#(r) \rightarrow v$: denotes the version $v$ of $L$ having  $LRoot$ $=$ $r$,\\		
	
	$\triangleright$ \textsc{Variables of TEE:} \\
	\hspace{1em} $SK_{\mathbb{E}}^{tee}, PK_{\mathbb{E}}^{tee}$: keypair of $\mathbb{E}$ under $\Sigma_{tee}$,\\	
	
	\hspace{1em} $SK_{\mathbb{E}}^{pb}, PK_{\mathbb{E}}^{pb}$: keypair of $\mathbb{E}$ under $\Sigma_{pb}$,\\
	\hspace{1em} $hdr_{last} \leftarrow \perp$: the last header created by $\mathbb{E}$,\\		
	\hspace{1em} $LRoot_{pb} \leftarrow \perp$: the last root of $L$ synced with PB,\\
	\hspace{1em} $LRoot_{cur} \leftarrow \perp$: the root of $L \cup blks_p$ (not synced with PB),\\	
	\hspace{1em} $ID_{cur} \leftarrow 1$: the current version of $L$ (not synced with PB),\\

	\smallskip
	$\triangleright$ \textsc{Declaration of functions:}
	
	\func{$Init$() \textbf{public}} {
		($SK_{\mathbb{E}}^{pb}$, $PK_{\mathbb{E}}^{pb}$)$ \leftarrow\Sigma_{pb}.Keygen()$;\\ 
		($SK_{\mathbb{E}}^{tee}$, $PK_{\mathbb{E}}^{tee}$)$ \leftarrow\Sigma_{tee}.Keygen()$;\\
		
		\textbf{Output}($PK_{\mathbb{E}}^{tee}, PK_{\mathbb{E}}^{pb}$); \\
	}					
	\smallskip
	
	\func{$Exec$($txs[], \partial st^{old}, ~\pi^{inc}_{next}, LRoot_{tmp}$) \textbf{public}}{
		
		\textbf{assert} $\partial st^{old}.root = hdr_{last}.stRoot$; \\
		
		$\partial st^{new}\!, rcs, txs_{er} \leftarrow~\!processTxs(txs, \partial st^{old}\!, \!\pi^{inc}_{next}, LRoot_{tmp})$;\\
		
		$\sigma \leftarrow \Sigma_{pb}.sign(SK_{\mathbb{E}}^{pb}, (LRoot_{pb}, LRoot_{cur}))$; \\
		\textbf{Output}($LRoot_{pb}, LRoot_{cur}, \partial st^{new}, hdr_{last}, rcs$, $txs_{er}$, $\sigma$); \\	
		
	}
	\smallskip
	
	\func{$Flush$() \textbf{public}}{									
		$LRoot_{pb} \leftarrow LRoot_{cur}$; \Comment{Shift the version of $L$ synchronized with PB.} \\					
	}
	\smallskip
	
	\func{$processTxs$($txs[], \partial st^{old}, ~\pi^{inc}_{next}, ~LRoot_{tmp}$) \textbf{private}}{
		
		$\partial st^{new}, rcs[], txs_{er} \leftarrow$ runVM($txs$, $\partial st^{old}$); \Comment{Run $txs$ in VM.} \\
		$txs \leftarrow txs \setminus txs_{er}$; \Comment{Filter out parsing errors/wrong signatures.} \\
		
		$hdr\!\leftarrow~\!\mathbf{Header}(ID_{cur}, MkRoot(txs),$ \\
		~~~~~~~~~~~~$MkRoot(rcs), \partial st^{new}\!\!.root))$;\\ 
		$hdr_{last} \leftarrow  hdr$; \\
		$ID_{cur} \leftarrow  ID_{cur} + 1$; \\		
		$LRoot_{cur} \leftarrow newLRoot(hdr, ~\pi^{inc}_{next}, ~LRoot_{tmp})$; \\
		\textbf{return} $\partial st^{new}$, $rcs$, $txs_{er}$; \\
	}
	\smallskip
	
	\func{$newLRoot(hdr, ~\pi^{inc}_{next}, LRoot_{tmp})$ \textbf{private}}{
		\Comment{A modification of the incr. proof. template to contain $hdr$ \hfill}\\
		\textbf{assert} $\#(LRoot_{cur}) + 1 = \#(LRoot_{tmp})$; \Comment{1 block $\Delta$.} \\
		\textbf{assert} $\pi^{inc}_{next}.Verify(LRoot_{cur},~LRoot_{tmp})$;\\
		
		$\pi^{inc}_{next}[\text{-}1] \leftarrow h(hdr)$; \\
		
		\textbf{return} $deriveNewRoot(\pi^{inc}_{next})$; \\
	}
	\smallskip
	
	\caption{\footnotesize The program $prog^{\mathbb{E}}$ of enclave $\mathbb{E}$}
	\label{alg:enclave-VM}
	\vspace{-0.2cm}
\end{algorithm}

\begin{algorithm}[t] 
	\caption{\scriptsize The program $prog^{\mathbb{S}}$ of the smart contract $\mathbb{S}$ }\label{alg:log-smart-contract}
	\scriptsize 
	
	\SetKwProg{func}{function}{}{}

	\smallskip
	$\triangleright$ \textsc{Declaration of types and constants:}\\		
	\hspace{1em} \textbf{CensInfo} \{ $etx, equery, status, edata$ \},  \\
	\hspace{1em} $msg$: a current transaction that called $\mathbb{S}$,  \\
	
	\smallskip
	$\triangleright$ \textsc{Declaration of functions:}
	
	\func{$Init$($PK_{\mathbb{E}}^{pb}, PK_{\mathbb{E}}^{tee}, PK_{\mathbb{O}} $) \textbf{public} }{
		$PK_{\mathbb{E}}^{tee}[].add(PK_{\mathbb{E}}^{tee})$; \Comment{PK of $\mathbb{E}$ under $\Sigma_{tee}$.} \\ 
		$PK_{\mathbb{E}}^{pb}[].add(PK_{\mathbb{E}}^{pb})$; \Comment{PK of $\mathbb{E}$ under $\Sigma_{pb}$.} \\
		$PK_{\mathbb{O}}^{pb} \leftarrow PK_{\mathbb{O}}$; \Comment{PK of $\mathbb{O}$ under $\Sigma_{pb}$.} \\
		$LRoot_{pb} \leftarrow \perp$; \Comment{The most recent root hash of $L$ synchronized with $\mathbb{S}$.} \\ 
		$censReqs \leftarrow []$; \Comment{Request that $\mathbb{C}$s wants to resolve publicly.} \\
	}
	
	\func{$PostLRoot$($root_A, root_B, \sigma$) \textbf{public} }{
		\Comment{Verify whether a state transition was made within $\mathbb{E}$. \hfill} \\
		\textbf{assert} $\Sigma_{pb}.verify((\sigma, PK_{\mathbb{E}}^{pb}[\text{-}1]), (root_A, root_B))$;  \\			
		
		\Comment{Verify whether a version transition extends the last one. \hfill} \\
		\If{$LRoot_{pb} = root_A$}{
			$LRoot_{pb} \leftarrow root_{B}$; \Comment{Do a version transition of L.} \\
		} 				
	}
	
	\func{$ReplaceEnc$($PKN_{\mathbb{E}}^{pb}, PKN_{\mathbb{E}}^{tee}, r_{A}, r_{B}, \sigma, \sigma_{msg}$) \textbf{public} }{
		\Comment{Called	 by $\mathbb{O}$ in the case of enclave failure.\hfill}\\
		
		\textbf{assert} $\Sigma_{pb}.verify((\sigma_{msg}, PK_{\mathbb{O}}^{pb}), msg)$; \\ $PostLRoot(r_{A}, r_{B}, \sigma)$ ; \Comment{Do a version transition.} \\

		$PK_{\mathbb{E}}^{tee}.add(PKN_{\mathbb{E}}^{tee})$; \Comment{Upon change, $\mathbb{C}s$ make remote attestation.} \\ 
		$PK_{\mathbb{E}}^{pb}.add(PKN_{\mathbb{E}}^{pb})$; \\	
	}
	
	\func{$SubmitCensTx$($etx, \sigma_{msg}$) \textbf{public} }{
		\Comment{Called by $\mathbb{C}$ in the case her TX is censored.\hfill \hfill \hfill}\\		
		accessControl($\sigma_{msg}, msg.PK_{\mathbb{C}}^{pb}$); \\

		$censReqs$.add(\textbf{CensInfo}($etx, \perp, \perp, \perp$)); \\			
		
	}
	\smallskip
	
	\func{$ResolveCensTx(idx_{req}, status, \sigma$) \textbf{public} }{
		\Comment{Called by $\mathbb{O}$ to prove that $\mathbb{C}$'s TX was processed.\hfill \hfill \hfill}\\
		
		\textbf{assert} $idx_{req} < |censReqs|$;\\
		$r \leftarrow censReqs[idx_{req}]$; \\
		
		\textbf{assert} $\Sigma_{pb}.verify((\sigma, PK_{\mathbb{E}}^{pb}[\text{-}1]), ~(h(r.etx), status))$; 	\\ 
		$r.status \leftarrow status$;\\
	}	
	
	\func{$SubmitCensQry$($equery, \sigma_{msg}$) \textbf{public} }{
		\Comment{Called by $\mathbb{C}$ in the case its read query is censored.\hfill \hfill}\\		
		accessControl($\sigma_{msg}, msg.PK_{\mathbb{C}}^{pb}$); \\

		$censReqs$.add(\textbf{CensInfo}($\perp, equery, \perp, \perp$)); \\		
		
	}
	\smallskip
	
	\func{$ResolveCensQry(idx_{req}, status, edata, \sigma$) \textbf{public} }{
		\Comment{Called by $\mathbb{O}$ as a response to the $\mathbb{C}$'s censored  read query.\hfill \hfill \hfill}\\		
		\textbf{assert} $idx_{req} < |censReqs|$;\\
		$r \leftarrow censReqs[idx_{req}]$; \\
		
		\textbf{assert}~$\Sigma_{pb}.verify((\sigma, PK_{\mathbb{E}}^{pb}[\text{-}1]),$ \\
		 ~~~~~~~~$(h(r.equery), status, h(edata)))$; \\ 
		$r.\{edata \leftarrow edata, status \leftarrow status\}$;\\		
	}
	\vspace{-0.1cm}
\end{algorithm}

\subsection{Normal Operation}\label{sec:normal-operation}
\begin{figure}[t]
		\vspace{-0.2cm}
	\begin{center}
		\includegraphics[width=0.46\textwidth]{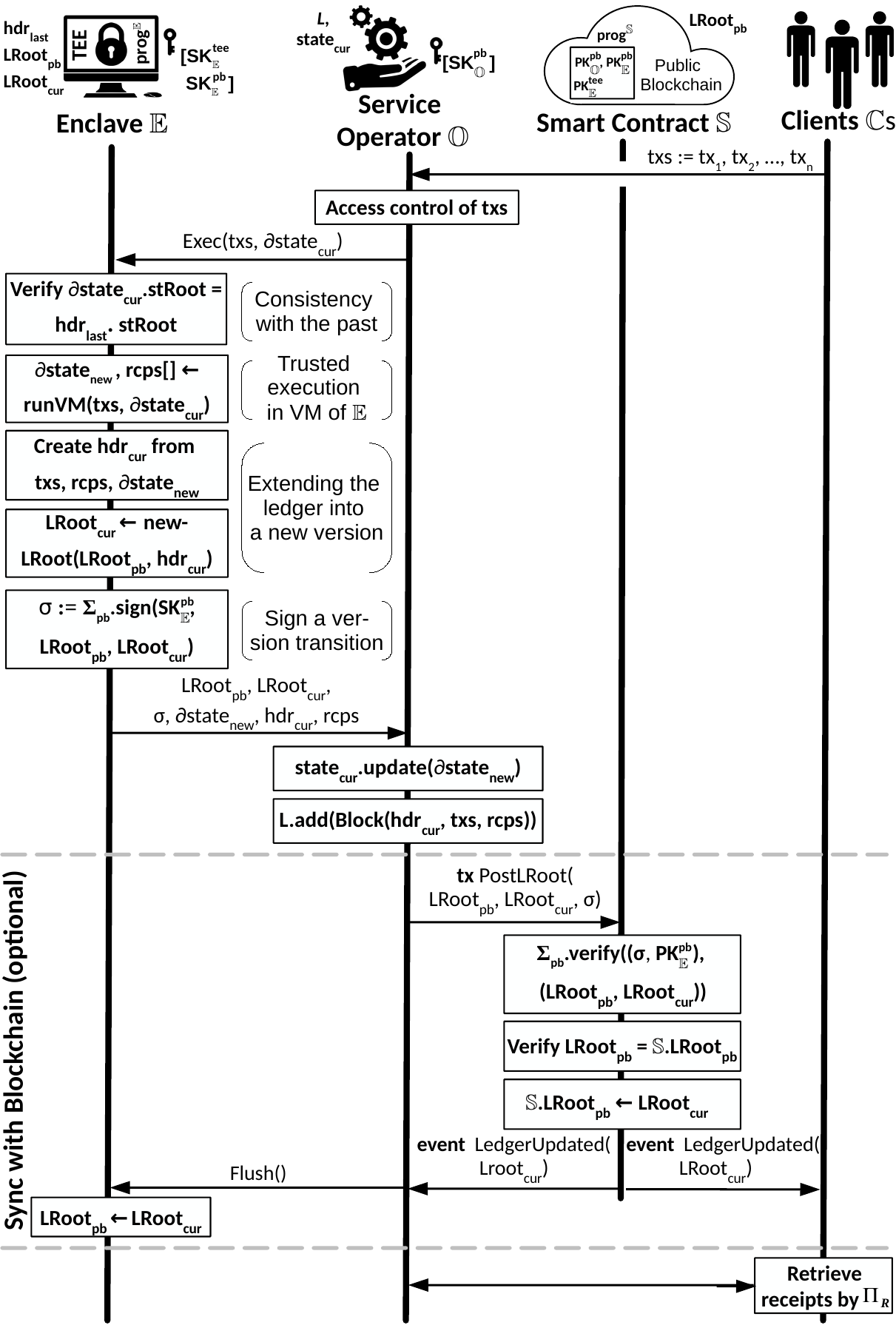} 
		\vspace{-0.1cm}
		\caption{Protocol for normal operation ($\Pi_{N}$).}		
		\label{fig:normal-operation}	
		\vspace{-0.7cm}	
	\end{center}	
\end{figure}

In the case of normal operation (see \autoref{fig:normal-operation}), $\mathbb{O}$ is not censoring any transactions produced by instances of $\mathbb{C}$, hence all transactions are correctly executed within $\mathbb{E}$ and appended to $L$, while $\mathbb{S}$ publicly witnesses the correct execution of transactions and the consistency of the new version of $L$ with its history.
In detail, when $\mathbb{O}$ receives a transaction from $\mathbb{C}_x$, it performs access control of $\mathbb{C}_x$,
and upon the success, $\mathbb{O}$ adds the transaction in its cache of unprocessed transactions (see \autoref{alg:operator} in Appendix).
When $\mathbb{O}$ accumulates enough transactions from clients, it passes these transactions to $\mathbb{E}$, together with the current partial state $\partial state_{cur}$ of the VM.

\subsubsection{\textbf{VM Execution with Partial State}}\label{sec:vm-execution-with-partial-state}
The current partial state of VM represents only data related to all account states that the execution of transactions is about to modify or create.
The motivation for such an approach is the limited memory size of $\mathbb{E}$ (e.g., in the case of SGX it is only $\sim$100MB), which does not allow for internally storing the full global state of $L$ (neither $L$ itself).
The partial state does not contain only the account states of concerning transactions, but it also contains intermediary nodes of MPT (i.e., extension and branch nodes) that are on the path from the root node of MPT to leaf nodes of concerning account states.
Using passed partial state, $\mathbb{E}$ verifies its integrity, obtains the state root of MPT, and compares it with the last known state root (i.e., $hdr_{last}.stRoot$) produced by $\mathbb{E}$.
If the roots match, $\mathbb{E}$ executes transactions using the passed partial state, and it obtains the new partial state of VM and execution receipts with additional information about the execution of particular transactions (i.e., return codes and logs).
Note that $\mathbb{E}$ obtains the new partial state by consecutively updating the current partial state with each transaction executed.

Next, $\mathbb{E}$ creates the header of the new block (i.e., $hdr_{cur}$) from aggregated transactions, receipts, and new partial state.
Using the created header,  
$\mathbb{E}$ extends the previous version of the history tree of $L$, while obtaining the new root hash $LRoot_{cur}$ of $L$ (see \autoref{sec:updating-the-ledger}).
Then, $\mathbb{E}$ signs a version transition pair $\langle LRoot_{pb}, LRoot_{cur} \rangle$ of the history tree by $SK_\mathbb{E}^{pb}$ and sends it to $\mathbb{O}$, together with the new header, the new partial state, and execution receipts. 
Moreover, $\mathbb{E}$ stores the last produced header (i.e., $hdr_{last}$) and the root hash $LRoot_{cur}$ of $L$ associated with the last version of the history tree.

\subsubsection{\textbf{Incremental Update of $L$}}\label{sec:updating-the-ledger}
We abstracted from the details about the consistent update of $L$ within $\mathbb{E}$ in the above text and \autoref{fig:normal-operation}.
In general, an incremental update of a history tree assumes trusted full access to its data.
However, $\mathbb{E}$ does not store full data of $L$ (only the last header created), and thus cannot directly make a consistent update of $L$.
Therefore, we design a procedure that enables $\mathbb{E}$ to extend $L$ without storing it: 
In detail, $\mathbb{O}$ creates a proof template $\pi^{inc}_{next}$ for the next incremental proof of the history tree, which extends $L$ exactly by one empty block (see $nextIncProof()$ in \autoref{alg:operator}) while obtaining a new version of $L$ with the (temporary) root hash $LRoot_{tmp}$.
Note that this template represents $\partial L$ from \autoref{sec:high-overview}, and it enables $\mathbb{E}$ to make an integrity verification and consistent extension of $L$ without storing it. 
In detail, $\mathbb{O}$ sends $\pi^{inc}_{next}$ and $LRoot_{tmp}$ to $\mathbb{E}$, together with transactions that are about to be processed by function $Exec()$. 
$\mathbb{E}$ verifies $\pi^{inc}_{next}$ with respect to its last known version of $L$ (i.e., $\#(LRoot_{cur})$), replaces the header hash of the empty block in the proof template by the hash of newly created header in $\mathbb{E}$ and then uses such modified proof to compute the new root hash of $L$, which is then stored as $LRoot_{cur}$ by $\mathbb{E}$.

When $\mathbb{O}$ receives the output of $\mathbb{E}$, it updates the full state of $L$ and creates the new block using client transactions, the received receipts, and the header of the new block.
Then, $\mathbb{O}$ appends the new block to $L$ and responds to client requests for receipts of their transactions (see \autoref{sec:receipt-retrieval}), which serve as \textit{promises} confirming the execution of transactions.
These promises became irreversible when $O$ syncs $L$ with $\mathbb{S}$.

\subsubsection{\textbf{Syncing the Ledger with the Blockchain}}
$\mathbb{O}$ periodically syncs $L$ with $\mathbb{S}$ to provide non-equivocation of $L$.
However, $\mathbb{O}$ can sync only such a version of $L$  that was signed within $\mathbb{E}$ and is newer than the last known version by $\mathbb{S}$, which provides consistency and non-equivocation of $L$.
During the sync of $L$, $\mathbb{O}$ creates a special blockchain transaction containing the version transition pair $\langle LRoot_{pb}, LRoot_{cur} \rangle$ signed within $\mathbb{E}$ and sends it to $\mathbb{S}$ (i.e., calling the function $PostLRoot()$).
$\mathbb{S}$ verifies whether the version transition pair was signed within  $\mathbb{E}$ by checking the signature with $PK_{\mathbb{E}}^{pb}$.
Then, $\mathbb{S}$ verifies whether the last published version of $L$ (corresponding to $LRoot_{pb}$ at $\mathbb{S}$) is equal to the first entry in the version transition pair.
If so, $\mathbb{S}$ publicly performs the version transition of $L$ by updating its $LRoot_{pb}$ to the second item of the version transition pair.
From that moment, the \name transactions processed until the current version of $L$ cannot be tampered with -- providing a non-equivocation of $L$.
Finally, $\mathbb{O}$ notifies $\mathbb{E}$ about successful sync by calling function $Flush()$ (see \autoref{alg:enclave-VM}), where $\mathbb{E}$ ``shifts'' $LRoot_{pb}$ to $LRoot_{cur}$. 

Note that if $\mathbb{O}$ were to sync $L$ with $\mathbb{S}$ upon every new block created, it might be expensive.
On the other hand, if $\mathbb{O}$ were to sync $L$ to $\mathbb{S}$ with long delays, ``a level'' of non-equivocation would be decreased, which in turn would extend the time to finality.
Hence, the sync interval must be viewed as a trade-off between costs and a level of non-equivocation (see examples in \autoref{sec:cost-analysis}).
The frequency of syncs might be defined in SLA with clients and violation might be penalized by $\mathbb{S}$.

\begin{figure}[t]
	\begin{center}
\includegraphics[width=0.4\textwidth]{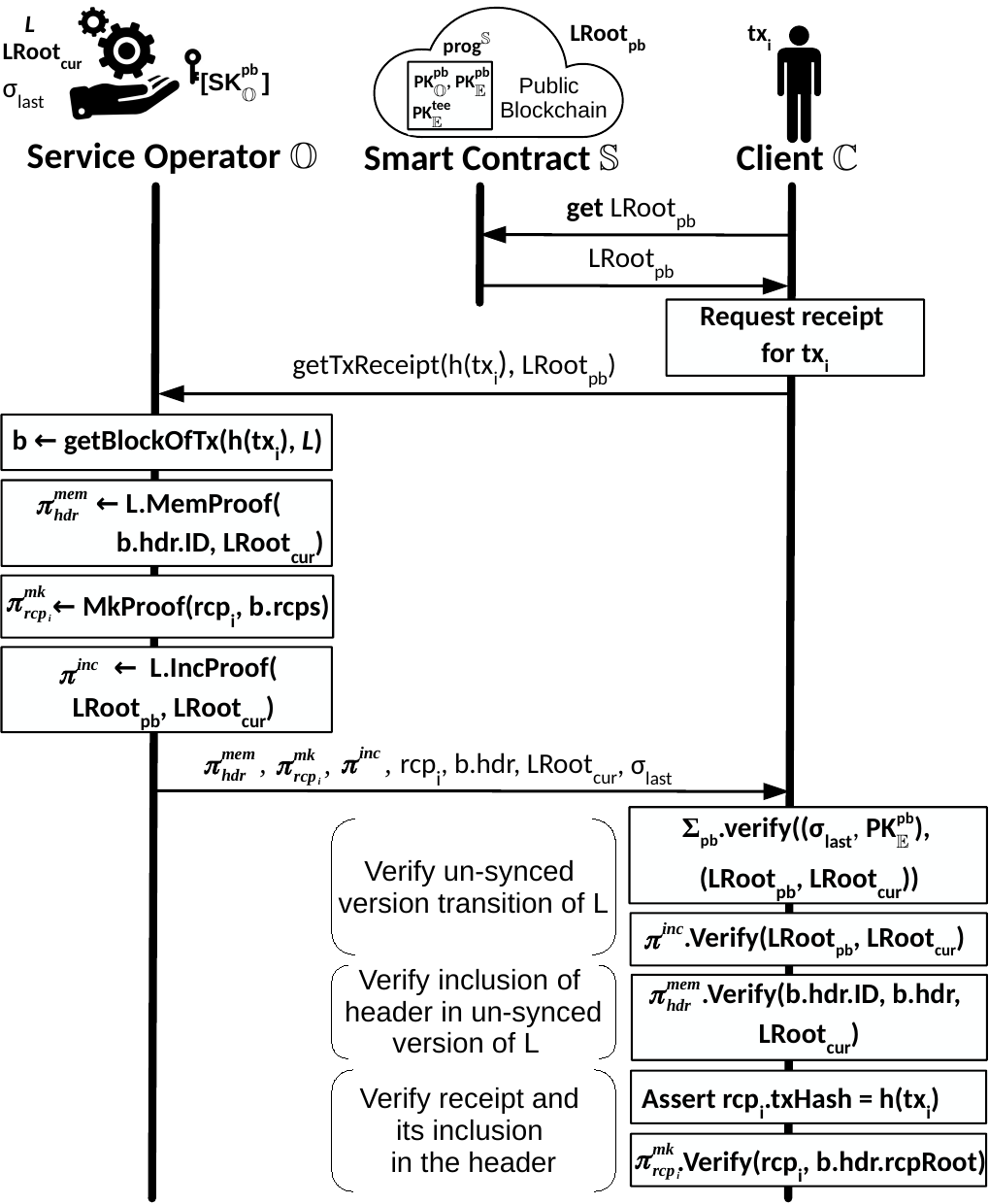} 
		\vspace{-0.1cm}
		\caption{Protocol for  receipt retrieval ($\Pi_{R}$).}		
		\label{fig:audit}
		\vspace{-0.6cm}
	\end{center}	
\end{figure}

\subsection{Retrieval and Verification of Receipts}\label{sec:receipt-retrieval}
Receipt retrieval and verification serve as a lightweight audit procedure in which $\mathbb{C}$ verifies the inclusion and execution of $tx_i$ by obtaining its receipt. 
An execution receipt contains three fields: (1) hash of $tx_i$, (2) return code of the VM, and (3) log of events emitted by the~VM. 

To obtain an execution receipt of $tx_i$, $\mathbb{C}$ first retrieves the last root hash of $L$ (i.e., $LRoot_{pb}$) from $\mathbb{S}$.
Then, $\mathbb{C}$ requests $\mathbb{O}$ for an inclusion proof of her transaction $tx_i$ in the most recent version of $L$ that extends the version $\#(LRoot_{pb})$.
Upon request, $\mathbb{O}$ finds a block $b$ that contains $tx_i$ and computes a membership proof $\pi^{mem}_{hdr}$ of $b$'s header in the most recent version $\#(LRoot_{cur})$ of $L$. 
The second proof that $\mathbb{O}$ computes is the Merkle proof $\pi^{mk}_{rcp_i}$, which witnesses that receipt $rcp_i$ of transaction $tx_i$ is included in the block $b$.
Then, $\mathbb{O}$ computes the incremental proof $\pi^{inc}$ of the most recent version transition $\langle  LRoot_{pb} ,LRoot_{cur} \rangle$ that was executed within $\mathbb{E}$. 
In response, $\mathbb{O}$ sends the following data to $\mathbb{C}$:
\begin{inparaenum}
	\item the receipt $rcp_i$ with its proof $\pi^{mk}_{rcp_i}$,
	\item the header of $b$ with its proof $\pi^{mem}_{hdr}$,
	\item the most recent version $LRoot_{cur}$ of $L$ with its proof $\pi^{inc}$,
	\item the signature $\sigma_{last}$ of the most recent version transition $\langle  LRoot_{pb} ,LRoot_{cur} \rangle$ made by $\mathbb{E}$.
\end{inparaenum}

$\mathbb{C}$ verifies the signature and the proofs against $LRoot_{pb}$, and it also checks whether the retrieved receipt corresponds to $tx_i$.
Then, $\mathbb{C}$ has a guarantee that the transaction $tx_i$ was included in $L$ and its execution in VM exited with a particular status, represented by a return code in the receipt.
We highlight that the previous receipt retrieval protocol assumes that $tx_i$ is ``very recent,'' and is included only in the version of $L$ that was not synchronized with $\mathbb{S}$ yet.
When $tx_i$ is already included in the synchronized version of $L$, we can put $LRoot_{cur} = LRoot_{pb}$, and thus skip computation of $\pi^{inc}$ and its verification.
We also note that the receipt retrieval protocol can be integrated with $\Pi_N$ by following it.

\subsection{Resolution of Censored Transactions}\label{sec:censored-transaction}
\begin{figure}[t]
	\begin{center}
		\centering 
				\vspace{-0.2cm}
		\includegraphics[width=0.48\textwidth]{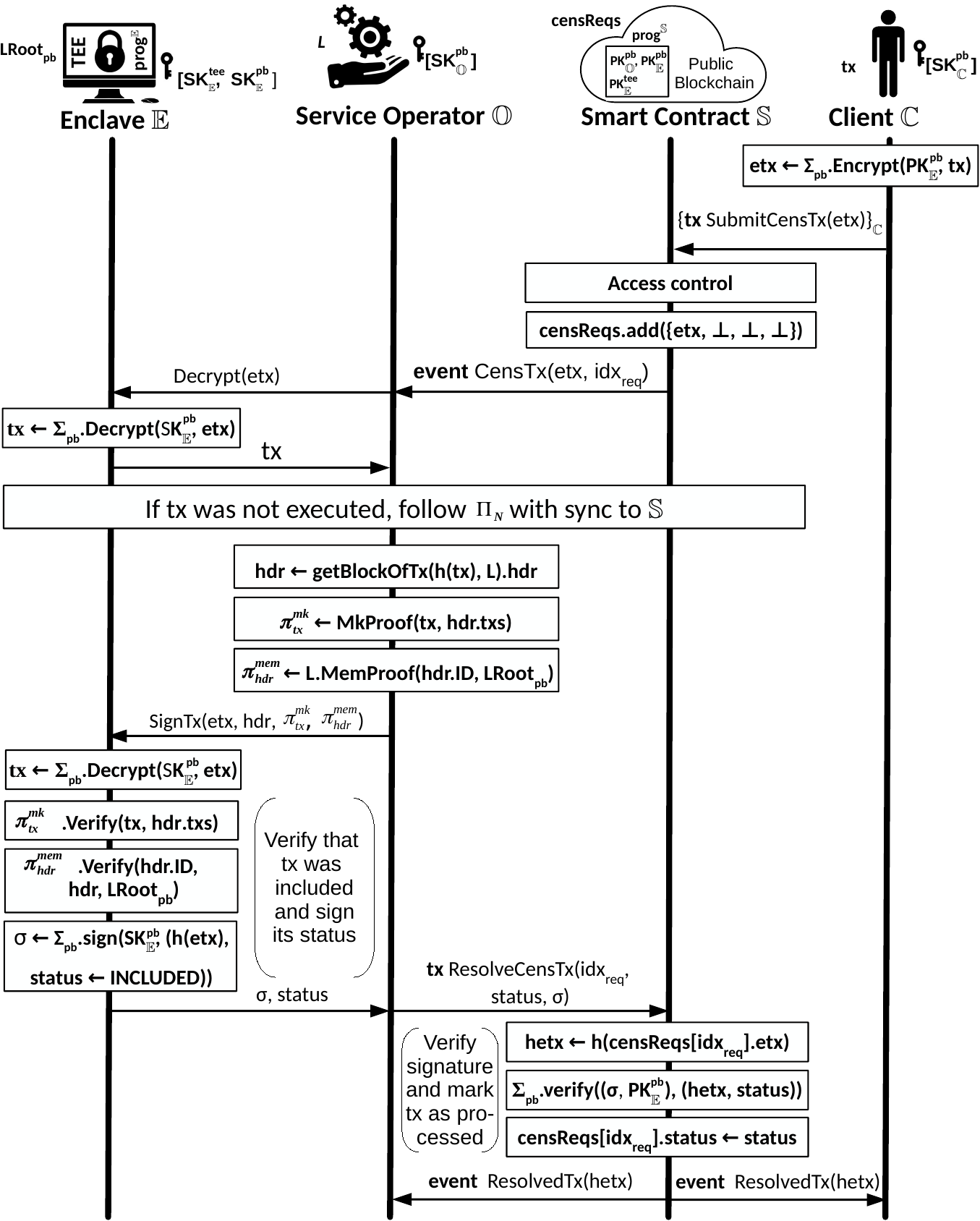} 
		\vspace{-0.4cm}
		\caption{Protocol for resolution of a censored tx ($\Pi_{CT}$).}		
		\label{fig:censored-tx}
		\vspace{-0.6cm}
	\end{center}	
\end{figure}

If $\mathbb{C}$ suspects $\mathbb{O}$ from censoring $tx$ (see \autoref{fig:censored-tx}), $\mathbb{C}$ initiates a request for an inclusion proof of $tx$ through $\mathbb{S}$.
In detail, $\mathbb{C}$ creates a transaction of the public blockchain, which calls the function $SubmitCensTx()$ with $tx$ encrypted by $PK_\mathbb{E}^{pb}$ (i.e., $etx$) as an argument and sends it to $\mathbb{S}$; hence, preserving confidentially for the public.
$\mathbb{S}$ does the access control (see \autoref{sec:cens-write-access}), appends $etx$ to the list of censored requests,\footnote{Note that to save costs for allocating storage of the smart contract platform, $\mathbb{S}$ can store only the hash of $etx$ instead (see \autoref{sec:exps-censorship}).} and generates asynchronous event informing $\mathbb{O}$ about new unresolved transaction.
When $\mathbb{O}$ receives the event, first she decrypts $tx$ through $\mathbb{E}$ and then executes $tx$ in $\mathbb{E}$ if it has not been executed before.
If a fresh execution of $tx$ occurred, $\mathbb{O}$ syncs the most recent version of $L$ with $\mathbb{S}$.
Then, $\mathbb{O}$ sends the encrypted $tx$ to $\mathbb{E}$ (i.e., function $SignTx()$) together with the header and the proofs that bind $tx$ to $L$, i.e., to its version $\#(LRoot_{pb})$.
In the function $SignTx()$ (see \autoref{alg:enclave-VM}), $\mathbb{E}$ decrypts $tx$ and checks whether it is correctly parsed and its signature is correct.
If these checks are not successful, $\mathbb{E}$ includes this information in the status of the response and signs it.
If the checks are successful, $\mathbb{E}$ proceeds to the verification of provided proofs with regard to the version $\#(LRoot_{pb})$ of $L$ synchronized to $\mathbb{S}$.
Upon successful verification, $\mathbb{E}$ signs both the transaction's status and the hash of encrypted $tx$, and then returns them to $\mathbb{O}$, who publishes the signature and the status through $\mathbb{S}$ (i.e., the function $ResolveCensTx()$).
When $\mathbb{S}$ receives the message with the status of $tx$ signed by $\mathbb{E}$, it computes the hash of $etx$ and uses it in the signature verification.
In the successful case, $\mathbb{S}$ updates the status of the censored transaction with the result from $\mathbb{E}$.
Finally, $\mathbb{S}$ notifies $\mathbb{C}$ and $\mathbb{O}$ about the resolution of $tx$.

\subsection{Resolution of Censored Queries}\label{sec:censored-query}
\begin{figure}[t]
	\begin{center}
		\centering
		\vspace{-0.2cm}
		\includegraphics[width=0.48\textwidth]{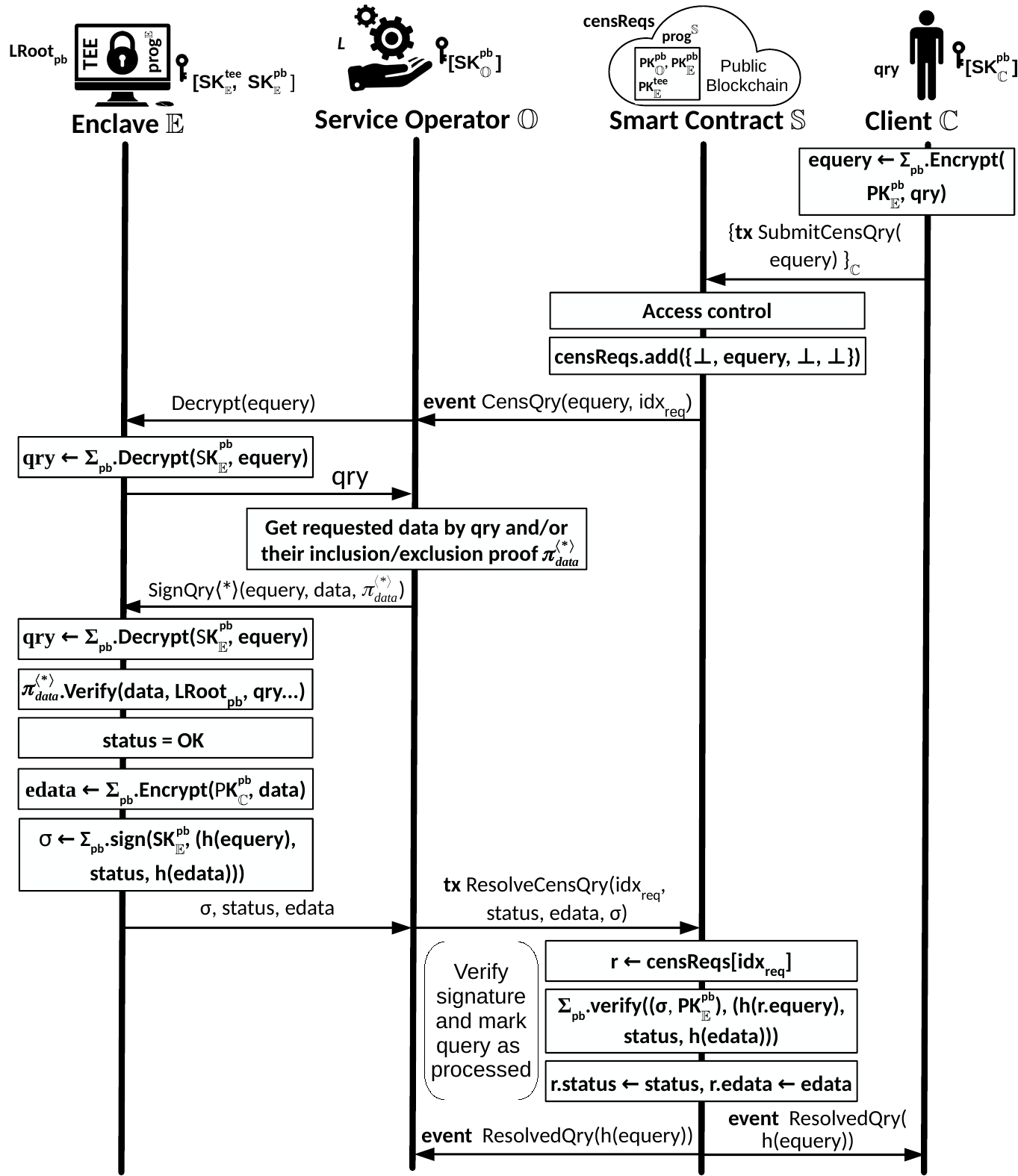}
		\vspace{-0.5cm} 
		\caption{Protocol for resolution of a censored query ($\Pi_{CQ}$).}		
		\label{fig:censored-query}
		\vspace{-0.6cm} 
	\end{center}	
\end{figure}

Besides censoring transactions, $\mathbb{O}$ might also censor read queries of $\mathbb{C}$s.
When $\mathbb{C}$ suspects $\mathbb{O}$ of censoring a read query $qry$ (see \autoref{fig:censored-query}), $\mathbb{C}$ initiates inclusion of $qry$ through $\mathbb{S}$.
In detail, $\mathbb{C}$ creates a transaction of the public blockchain, which calls the function $SubmitCensQry()$ (see \autoref{alg:log-smart-contract}) with $qry$ encrypted by $PK_\mathbb{E}^{pb}$ (i.e., $equery$) as an argument and sends it to $\mathbb{S}$.
$\mathbb{S}$ does the access control (see \autoref{sec:cens-write-access}), appends $equery$ to the list of censored requests, and generates an event informing $\mathbb{O}$ about the new unresolved query.
When $\mathbb{O}$ receives the event, first she decrypts $qry$ through $\mathbb{E}$, then fetches the data requested by $qry$ and computes their inclusion proof(s) $\pi_{data}^{<*>}$ in the version of $L$ synchronized to $\mathbb{S}$.\footnote{$\mathbb{O}$ creates an exclusion proof when requesting non-existing data.}
Then, $\mathbb{O}$ sends the encrypted $qry$ to $\mathbb{E}$ (i.e., function $SignQry\langle*\rangle()$) together with the fetched data and their inclusion proof $\pi_{data}^{<*>}$.
In the function $SignQry\langle*\rangle()$, $\mathbb{E}$ decrypts $qry$ and checks whether it is correctly parsed.
Next, $\mathbb{E}$ proceeds to the verification of $\pi_{data}^{<*>}$ with regard to the version $\#(LRoot_{pb})$ of $L$ synchronized with $\mathbb{S}$.
Upon successful verification, $\mathbb{E}$ encrypts data by $PK_{\mathbb{C}}^{pb}$ (i.e., $edata$) and signs the triplet consisting of the transaction's status, the hash of $eqry$, and the hash of $edata$; which are returned to $\mathbb{O}$.
Then, $\mathbb{O}$ calls the function $ResolveCensQry()$ of $\mathbb{S}$ (see \autoref{alg:log-smart-contract}) with signature, status, and encrypted data contained in the arguments.
When $\mathbb{S}$ receives the message with the status of $qry$ signed by $\mathbb{E}$, it computes the hashes of $equery$ and $edata$, and it uses them in the signature verification.
Upon successful verification, $\mathbb{S}$ updates the status and $edata$ of the suspected censored query with the data from $\mathbb{E}$.
Finally, $\mathbb{S}$ notifies $\mathbb{C}$ and $\mathbb{O}$ about the resolution of $qry$.
We provide related pseudo code of $\mathbb{E}$ and examples of handling different queries in Appendix~\ref{sec:censored-query-example}.

\subsection{Terminated and Failed Enclave}\label{sec:failed-enclave}
During the execution of $prog^\mathbb{E}$, $\mathbb{E}$ stores its secrets and state objects in a sealed file, which is updated on the hard drive of $\mathbb{O}$ with each new block.
If $\mathbb{E}$ terminates due to a temporary reason, such as a power outage or command by $\mathbb{O}$, it can be initialized again by $\mathbb{O}$ who provides $\mathbb{E}$ with the sealed file used for recovery of the protected state. 
Note that if sealed file were to be tampered with by malicious $\mathbb{O}$ through the rollback attack~\cite{matetic2017rote} reverting the state of $L$ to inconsistent version with $\mathbb{S}$, $\mathbb{E}$ would not be able to create valid state transition pairs that could be accepted by $\mathbb{S}$ due to version mismatch (i.e., ($\#(\mathbb{S}.LRoot_{pb}) \neq \#(\mathbb{E}.LRoot_{pb}$)).

However, if $\mathbb{E}$ experiences a permanent hardware failure of TEE, the sealed file cannot be decrypted on other TEE platforms.
Therefore, we propose a simple mechanism that deals with this situation under the assumption that $\mathbb{O}$ is the only allowed entity that can replace the platform of $\mathbb{E}$.
In detail, $\mathbb{O}$ first snapshots the header $hdr_{sync}$ of the last block that was synchronized with $\mathbb{S}$ as well as all blocks $blks_{unsync}$ of $L$ that were not synchronized with $\mathbb{S}$.
Then, $\mathbb{O}$ restores $L$ and her  internal state objects into the version $\#(LRoot_{pb})$.
After the restoration of $L$, $\mathbb{O}$ calls the function $ReInit()$ of $\mathbb{E}$ (see \autoref{alg:enclave-VM-failure}) with $hdr_{sync}$, $blks_{unsync}$, and $LRoot_{pb}$ as the arguments.
In this function, $\mathbb{E}$ first generates its public/private key-pair $SK_\mathbb{E}^{pb}, PK_\mathbb{E}^{pb}$, and then stores the passed header as $hdr_{last}$ and copies the passed root hash into $LRoot_{cur}$ and $LRoot_{pb}$.
Then, $\mathbb{E}$ iterates over all passed unprocessed blocks and their transactions $txs$, which are executed within the VM of $\mathbb{E}$.
Before the processing of $txs$ of each passed block, $\mathbb{E}$ calls the unprotected code of $\mathbb{O}$ to obtain the current partial state $\partial st^{old}$ of $L$ and incremental proof template (see \autoref{sec:updating-the-ledger}) that serves for extending $L$ within $\mathbb{E}$.
However, these unprotected calls are always verified within $\mathbb{E}$, and malicious $\mathbb{O}$ cannot misuse them.
In detail, $\mathbb{E}$ verifies $\partial st^{old}$ obtained from $\mathbb{O}$ against the root hash of the state stored in the last header $hdr_{last}$ of $\mathbb{E}$, while the incremental proof template is also verified against $LRoot_{cur}$ in the function $newLRoot()$.

\begin{algorithm}[t] 	
	\scriptsize
	\SetKwProg{func}{function}{}{}
	
	\func{$ReInit$($LRoot_{old}$, $prevBlks[], hdr_{last}$) \textbf{public}} {
		
		($SK_{\mathbb{E}}^{pb}$, $PK_{\mathbb{E}}^{pb}$)  $ \leftarrow\Sigma_{pb}.Keygen()$;\\	
		$hdr_{last} \leftarrow hdr_{last}$; \\						
		$LRoot_{cur}  \leftarrow LRoot_{old}, LRoot_{pb} \leftarrow LRoot_{old}$;\\			
		\smallskip
		
		\For{$\{ b: prevBlks \}$}{	
			$\pi^{inc}_{next}, ~LRoot_{tmp} \leftarrow prog^{\mathbb{O}}.nextIncProof()$; \\
			$\partial st^{old} \leftarrow prog^{\mathbb{O}}.getPartialState(b.txs)$;\\
			\textbf{assert} $\partial st^{old}.root = hdr_{last}.stRoot$; \\
			
			$\ldots \leftarrow ~processTxs(b.txs, \partial st^{old}, \pi^{inc}_{next}, ~LRoot_{tmp})$;\\
			
			$LRoot_{ret} \leftarrow prog^{\mathbb{O}}.runVM(b.txs)$; \Comment{Run VM at $\mathbb{O}$.} \\
			\textbf{assert} $LRoot_{cur} = LRoot_{ret}$; \Comment{$\mathbb{E}$ and $\mathbb{O}$  have the same $L$.}\\
		}
		
		$\sigma \leftarrow \Sigma_{pb}.sign(SK_{\mathbb{E}}^{pb}, (LRoot_{pb}, LRoot_{cur}))$; \\
		
		\textbf{Output}($LRoot_{pb}, LRoot_{cur}, \sigma, PK^{\mathbb{E}}_{pb}, PK^{\mathbb{E}}_{tee}$); 	\\
		
	}					
	\smallskip

	\caption{\scriptsize Reinitialization of a failed $\mathbb{E}$ (part of $prog^{\mathbb{E}}$).}
	\label{alg:enclave-VM-failure}
	\vspace{-0.2cm}
\end{algorithm}

Next, $\mathbb{E}$ processes $txs$ of a block, extends $L$ and calls the unprotected code of $\mathbb{O}$ again, but this time to process $txs$ of the current block by $\mathbb{O}$, and thus getting the same version and state of $L$ in both $\mathbb{E}$ and $\mathbb{O}$. 
Note that any adversarial effect of this unprotected call is eliminated by the checks made after the former two unprotected calls.
When all passed blocks are processed, $\mathbb{E}$ signs the version transition pair $\langle LRoot_{pb}, LRoot_{cur} \rangle$ and returns it to $\mathbb{O}$, together with the new public keys of $\mathbb{E}$.
$\mathbb{O}$ creates a blockchain transaction that calls the function $ReplaceEnc()$ of $\mathbb{S}$ with data from $\mathbb{E}$ passed. In $ReplaceEnc()$, $\mathbb{S}$ first verifies whether the signature of the transaction was made by $\mathbb{O}$ to avoid MiTM attacks. Then, $\mathbb{S}$ calls its function $PostLRoot()$ with the signed version transition pair in the arguments.
Upon the success, the current root hash of $L$ is updated and $\mathbb{S}$ replaces the stored $\mathbb{E}$'s PKs by PKs passed in parameters.
Finally, $\mathbb{E}$ informs $\mathbb{C}s$ by an event containing new PKs of $\mathbb{E}$, and $\mathbb{C}$s perform the remote attestation of $prog^{\mathbb{E}}$ using the new key $PK^{\mathbb{E}}_{tee}$ and the attestation service.
We refer the reader to \autoref{alg:operator} in the Appendix for the pseudo-code of~$\mathbb{O}$.

\section{Universal Composability}
We model the execution of a protocol $\proto$ via the Universal Composability (UC) Framework by Canetti et al.~\cite{TCC:CDPW07} where all the entities are Probabilistic Polynomial-Time (PPT) Interactive Turing Machines and the execution is controlled by the environment $\env$.  In this simulation-based argument, all parties from $\proto$ perform the protocol, \ie the real execution $\realmodel$, in the presence of the adversary $\adv$ which can see and delay 
the messages. Whereas the ideal execution,  \ie $\idealmodel$, is composed by  the functionality $\funcmodel$ in the presence of the simulator $\IAdv$, and it is also controlled by $\env$. 
Given the security parameter $\secpar$, randomness $\rand$ and input $\envinput$, 
the $\zenv$ drives both executions $\idealmodel$ and $\realmodel$, and outputs either $1$ or $0$. Therefore, let $\idealvar$ and $\realmodel$ be the ensembles, \ie ideal $\ensideal$ and real $\ensreal$  of the outputs of $\zenv$ for both executions. 
Thus, we say that \textit{$\proto$ realizes $\funcmodel$} when there exist a PPT simulator $\IAdv$, such that for all PPT $\zenv$, we have $\idealvar\approx_c \realvar$.

\subsection{UC Security Definition}
We introduce \name Ideal Functionality $\AQledger$ and prove that the \name protocol $\proto_{AQ}$ UC realizes $\AQledger$ under the security assumptions of the cryptographic primitives used, assuming the adversary $\mathcal{A}$ who can corrupt $\mathbb{O}$ but not $\mathbb{E}$.

\subsubsection{\textbf{Intuition}}
We need to ensure that once $\mathbb{C}$ has submitted a transaction, she can verify its processing by the functionality of the receipt. 
With receipt,  $\mathbb{C}$ can verify whether the transaction was processed or detect whether $\mathbb{O}$ is withholding~it.
Moreover, to promote public verifiability, we allow $\mathbb{C}$ to ask for all the unhandled receipts, which is feasible given the access to a public blockchain.
A receipt ties a transaction and the state, \ie a version of $\ledger$, which $\mathbb{C}$ wants to verify whether her transaction was included in. 
In the case of $\mathbb{C}$ suspecting $\mathbb{O}$ from censoring a transaction, $\mathbb{C}$ performs a receipt verification step via receipt functionality, and $\mathbb{O}$ must handle this request. Otherwise, it would become publicly visible that she is censoring. 
Therefore, an interface that reliably publishes the list of handled requests (\ie receipt retrieval and resolution of queries) works as a publicly available audit procedure. Note that ``to verify the receipt'' in \autoref{fig:aqledger1}, the functionality checks directly if the transaction was indeed added to $\ledger$, which is the meaning of a valid receipt. 
If verification of the receipt fails, the transaction was not added to the current state of the ledger, \ie it was explicitly censored, and therefore the receipt returned is invalid.
In the case that $\mathbb{O}$ is corrupted, she does no action to handle the receipt. We remark that the design of the functionality assumes that on each attempt of transaction addition to $\ledger$ (regardless of whether it was censored), the functionality generates a receipt, therefore we do not need to specify an interface for the ``receipt issuing'' procedure.

\subsection{\name Ledger Ideal Functionality}\label{sec:aqledger}
We present functionality $\AQledger$ in \autoref{fig:aqledger1} and \autoref{fig:aqledger2}, representing another way of specifying the properties from \autoref{sec:desired-properties}.

\myhalfbox{\name Ledger $\AQledger$ (Part 1)}{white!40}{white!10}{
	\scriptsize
	
	\vspace{-0.2cm}
	\begin{justify}    
		The functionality interacts with an ideal adversary $\IAdv$ 
		and a set of $\client$s by replying to $\client_i$ and $\op$. 
		It keeps: 
		\begin{compactitem}
			\item[--]  $\submTxSet$, $\censoredTx$: sets of submitted and censored txs, 
			\item[--] $\submRecSet$, $\submRecNotHndSet$: sets of submitted and not handled receipts, 
			\item[--] $\submQrySet, \censQrySet, \submQryNotHndSet$: 
			sets of submitted/censored/not handled queries, 
			
			\item[--] $\ledgerStateList$ and $\ledgerStateListPub$: ordered lists of ledger states $\ledgerState_{i}$ (internal and public). 
			Each $\ledgerState_{i}$ keeps an ordered list of txs,
			
			\item[--] $\opCorrupt$: boolean variable indicating $\mathbb{O}$ corruption, 
			\item[--] $\batch$: a list of txs to be processed. 
		\end{compactitem}
		
\end{justify}
	
	\noindent {\bf $ \blacktriangleright$ \underline{Initialization:}}
	Upon receiving $\msg{Init}{}$ from  $\op$, set $\submTxSet$, $\censoredTx$, $\submRecSet$, $\submRecNotHndSet$, $\submQrySet$ and $\censQrySet$ to $\emptyset$. Set $\opCorrupt$ to $0$, $\ledgerState_{0}$ to $\emptyset$, and add $\ledgerState_{0}$ to $\ledgerStateList$ and $\ledgerStateListPub$. Return $\msg{InitOK}{}$ to $\op$.

	\smallskip
	\noindent {\bf $ \blacktriangleright$ \underline{Read Public Ledger:}} 
	On receiving $\msg{read}{}$ from  $\client_i$ or $\op$, return $\msg{readOK}{\ledgerStateListPub}$.     
	
	\smallskip
	\noindent {\bf $ \blacktriangleright$ \underline{Add Transaction:}} 
	Upon receiving $\msg{Add}{\tx}$ from  $\client_i$, perform:
	\begin{compactitem}    	
		\item Add $\tx$ to $\submTxSet$.
		\item Send $\msg{add}{\tx,\client_i}$ to $\IAdv$. \item Create and set the variables $\proc\leftarrow 1$ and $\mathsf{msg}\leftarrow\emptyset$.
		\item If $\opCorrupt=1$, then
		\begin{compactitem}			
			\item Wait for $\msg{add}{\action,\mathsf{msg}^\prime}$ from $\IAdv$.
			\item Set $\proc\leftarrow\action$ and $\mathsf{msg}\leftarrow\mathsf{msg}^\prime$.
		\end{compactitem}
		\item If $\proc=1$, then add $\tx$ to $\batch$.

		Otherwise, add $\tx$ to $\censoredTx$.
		Finally, return $\msg{AddOk}{\tx}$ to  $\client_i$.
	\end{compactitem}

	\smallskip
	\noindent {\bf $\blacktriangleright$ \underline{Publish Batch:}}
	Upon receiving $\msg{publish}{}$ from  $\op$, perform \textbf{Publish-Procedure}, then reply $\msg{publishOk}{}$ to $\op$.

	\smallskip
	\noindent {\bf $ \blacktriangleright$ \underline{Publish-Procedure:}}
	Send $\msg{publish}{\batch}$ to $\IAdv$.

	If $\opCorrupt=1$, then: 
\begin{compactitem}			
		\item Wait for the reply $\msg{publish}{\batch^\prime}$ from $\IAdv$. 
		
		\item $\censoredTx \leftarrow$ $\censoredTx ~\union$ $\forall\tx$ that represent $\batch\setminus\batch^\prime$. 
		
		\item $\batch\leftarrow\batch^\prime$.
	\end{compactitem}
	
	Given $\ledgerStateList$, for some $n$ such that $\ledgerStateList=(\ledgerState_{0}, \dots, \ledgerState_{n})$, do: 
	\begin{compactitem}				
		\item Create and set a list $\ledgerState_{n+1}\leftarrow\batch$.
		
		\item Add $\ledgerState_{n+1}$ to $\ledgerStateList$ and $\ledgerStateListPub$.
\item $\batch\leftarrow\emptyset$.
		
	\end{compactitem}
}{\label{fig:aqledger1} The \name Ledger Ideal Functionality (Part I).}

\subsubsection{\textbf{Corruption}} Functionality of the \name ledger $\AQledger$ allows the corruption of  $\op$. We assume that at any moment the simulator $\IAdv$ can submit a message $\msg{Corrupt}{}$\footnote{Note that $sid$ represents the session ID of the instance of $\proto_{AQ}$.} which would set an internal variable $\opCorrupt=1$. This variable is kept for clarity in the $\AQledger$ definition, and it drives the behavior of $\AQledger$ in the case that $\adv$ corrupts 	$\op$ in the real execution of the simulation. 
However, corrupted  $\op$ cannot forge signatures of $\mathbb{E}$.

\subsection{Security}
We construct a simulator $\IAdv$, simulating the real execution of $\proto_{AQ}$ to $\adv$. The view of the real execution, provided by $\IAdv$ to $\adv$  cannot be distinguishable to a PPT $\env$  from the ideal execution (\ie execution between $\AQledger$ and~$\IAdv$).

\begin{theorem}
Assuming a collision-resistant hash function, security of Merkle, incremental and membership proofs, and the Existential Unforgeability under Chosen Message Attack (EUF-CMA) scheme for $\mathbb{E}$, then $\proto_{AQ}$ realizes $\AQledger$ in the presence of $\adv$ who corrupts $\op$ but not $\mathbb{E}$. 
\end{theorem}    
\noindent
We refer the reader to Appendix~\ref{appendix:uc-proof} for the proof.

\myhalfbox{\name Ledger $\AQledger$  (Part 2)}{white!40}{white!10}{
	\scriptsize    
	
	\noindent {\bf $\blacktriangleright$ \underline{Resolve Censored Tx:}}
	Upon receiving $\msg{receipt}{\tx}$ from  $\client_i$:
\begin{compactitem}			
		\item If $\tx\notin\submTxSet$, then halt.
\item Otherwise, 
		\begin{compactitem}	
			\item Add $\tx$ to $\submRecSet$, and send  $\msg{receipt}{\tx,\client_i}$ to $\IAdv$.

			\item Create and set the boolean variable $\handle\leftarrow 1$.
		\end{compactitem}	
	\end{compactitem}

	If $\opCorrupt=1$, then:
	\begin{compactitem}			
		\item Wait for the reply $\msg{receipt}{\action}$ from $\IAdv$.
\item $\handle\leftarrow\action$.
	\end{compactitem}
	
	If $\handle=1$, then:  \begin{compactitem}			
		\item If $\exists\tx\in\batch$, then  do \textbf{Publish-Procedure}.   \Comment{Delayed handling}
		\item If $\exists\ledgerState_{j}\in\ledgerStateList$ s.t. $\tx\in\ledgerState_{j}$, then		
		\begin{compactitem}
			\item Return $\msg{receiptOK}{\mbox{``Receipt valid''}}$.   \Comment{Tx executed}
		\end{compactitem}
\item Else if $\exists\tx\in\censoredTx$ 
		\begin{compactitem}
			\item Remove $\tx$ from $\censoredTx$ and add it to $\batch$.
			\item Perform \textbf{Publish-Procedure}. 
\item Return $\msg{receiptOK}{\mbox{``Receipt valid''}}$.    \Comment{Tx resolved}
\end{compactitem}  
	\end{compactitem} 
Otherwise, \Comment{Operator corrupted, proceed = 0}

	\begin{compactitem}			
		\item Add $\tx$ to $\submRecNotHndSet$.     \Comment{Receipt handling ignored}
		\item Return $\msg{receiptOK}{\mbox{``Receipt invalid''}}$. 
	\end{compactitem}

	\smallskip
	\noindent {\bf $\blacktriangleright$ \underline{List Handled Receipts:}} 
	On receiving $\msg{handledReceipts}{}$ from  $\client_i$, reply $\msg{handledReceiptsOK}{ \submRecSet \setminus (\submRecNotHndSet \union \censoredTx )}$.

	\smallskip     
	\noindent {\bf $\blacktriangleright$ \underline{Query Ledger:}} 
	On receiving $\msg{Query}{\qry}$ from  $\client_i$, add $\qry$ to $\submQrySet$. Create the  variable $\mathsf{result}$, and if $\opCorrupt=0$, then 
	\begin{compactitem}	
		\item $\mathsf{result}\leftarrow\qry(\ledgerStateList)$. 
		\item Send $\msg{QueryOK}{\mathsf{result},\client_i}$ to $\IAdv$. 
	\end{compactitem}
	Else 
	\begin{compactitem}	
		\item Send $\msg{Query}{\qry,\client_i}$ to $\IAdv$ and wait for  $\msg{Query}{\mathsf{result}^\prime\!,\client_i\!}$. 
		\item $\mathsf{result}\leftarrow\mathsf{result}^\prime$.
	\end{compactitem}
	If  $\mathsf{result}\neq\qry(\ledgerStateList)$, add $\qry$ to $\censQrySet$. Return $\msg{QueryOK}{\mathsf{result}}$.

	\smallskip
	\noindent {\bf $\blacktriangleright$ \underline{List Resolved Queries:}}  On receiving $\msg{ResolvedQueries}{}$ from  $\client_i$, reply $\msg{ResolvedQueriesOK}{ \submQrySet \setminus  (\submQryNotHndSet \union\censQrySet)}$.

	\smallskip	
	\noindent {\bf $\blacktriangleright$ \underline{Resolve Censored Query:}} On receiving $\msg{ResolveQuery}{\qry}$ from  $\client_i$, forward $\msg{ResolveQuery}{\qry,\client_i}$ to $\IAdv$. Set $\handle\leftarrow1$, $\mathsf{msg}\leftarrow\emptyset$, $\mathsf{result}\leftarrow\qry(\ledgerStateList)$.  If $\opCorrupt=1$, then 	
	\begin{compactitem}
		\item Wait for $\IAdv$'s reply $\msg{ResolveQuery}{\action,\mathsf{result}^\prime,\mathsf{msg}^\prime}$.
		\item  Set $\handle\leftarrow\action$, $\mathsf{msg}\leftarrow\mathsf{msg}^\prime$ and $\mathsf{result}\leftarrow\mathsf{result}^\prime$.
	\end{compactitem}
	If $\handle=1$, then 
	\begin{compactitem}
		\item If $\qry\in\censQrySet$, remove it from $\censQrySet$ and add it to $\submQrySet$. 
		\item Otherwise,  add $\qry$ to  $\submQrySet$ and send $\msg{ResolveQueryOK}{ \mathsf{result},\mathsf{msg}}$ to $\client_i$. \end{compactitem}	
	Otherwise,  \Comment{Operator corrupted, proceed = 0}
	
	\begin{compactitem}			
		\item Add $\tx$ to $\submQryNotHndSet$ and halt.     \Comment{Query resolution ignored}
	\end{compactitem}

}{\label{fig:aqledger2} The \name Ledger Ideal Functionality (Part II).}

\section{Security Analysis and Discussion}
\label{sec:analysis}
On top of UC modeling, we describe (informally) resilience of \name against adversarial actions that $\mathcal{A}$ controlling $\mathbb{O}$ can perform to violate the desired properties (see \autoref{sec:desired-properties}). 
Then, we discuss other related aspects.

\begin{theorem}\label{theorem:correctness}
	(Correctness) $\mathcal{A}$ is unable to modify the state of $L$ while not respecting the semantics of VM in $\mathbb{E}$.
\end{theorem}
\begin{proof1}
	The update of the $L$'s state is made only in $\mathbb{E}$.	
	Since $\mathbb{E}$ contains trusted code that is publicly known and remotely attested by $\mathbb{C}$s, $\mathcal{A}$ cannot tamper with it.
\end{proof1}

\begin{theorem}\label{theorem:consistency}
	(Consistency) $\mathcal{A}$ is unable to extend $L$ while modifying the past records of $L$.
\end{theorem}
\begin{proof1}
	All extensions of $L$ are made in trusted code of $\mathbb{E}$, utilizing the history tree~\cite{crosby2009efficient} that enables us to make only such extensions of $L$ that are consistent with $L$'s past. 
\end{proof1}

\begin{theorem}
	(Verifiability) $\mathcal{A}$ is unable to unnoticeably modify or delete a transaction $tx$ that was previously inserted to $L$ using $\Pi_{N}$, if sync with $\mathbb{S}$ was executed anytime afterward.
\end{theorem}
\begin{proof1}
	Since $tx$ is correctly executed (\autoref{theorem:correctness}) in the block $b_i$ by trusted code of $\mathbb{E}$, $\mathbb{E}$ produces a signed version transition pair $\{h(L_{i-1}), h(L_i)\}_\mathbb{E}$ of $L$ to the new version $i$ that corresponds to $L$ with $b_i$ included.
	$\mathcal{A}$ could either sync $L$ with $\mathbb{S}$ immediately after $b_i$ is appended or she could do it $n$ versions later.
	In the first case, $\mathcal{A}$ publishes $\{h(L_{i-1}), h(L_i)\}_\mathbb{E}$ to $\mathbb{S}$, which updates its current version of $L$ to $i$ by storing $h(L_i)$ into $LRoot_{pb}$. 
	In the second case, $n$ blocks are appended to $L$, obtaining its $(i+n)$th version. 
	$\mathbb{E}$ executes all transactions from versions $(i+1),\ldots, (i+n)$  of $L$, while preserving correctness (\autoref{theorem:correctness}) and consistency (\autoref{theorem:consistency}). 
	Then, $\mathbb{E}$ generates a version transition pair $\{h(L_{i-1}), h(L_{i+n})\}_\mathbb{E}$ and $\mathcal{A}$  posts  it to $\mathbb{S}$, where 
	the current version of $L$ is updated to $i+n$ by storing $h(L_{i+n})$ into $LRoot_{pb}$.
	If any $\mathbb{C}$ requests $tx$ and its proofs from $\mathcal{A}$ with regard to publicly visible $LRoot_{pb}$, she might obtain a modified $tx'$ with a valid membership proof $\pi^{mem}_{hdr_i}$ of the block $b_i$ but an invalid Merkle proof  $\pi^{mk}_{tx'}$, which cannot be forged. 

	In the case of $tx$ deletion, $\mathcal{A}$ provides $\mathbb{C}$ with the tampered block $b_i'$ (maliciously excluding $tx$) whose membership proof $\pi^{mem}_{hdr_i'}$ is invalid -- it cannot be forged. 	
\end{proof1}

\begin{theorem}\label{theorem:non-equivocation}
	(Non-Equivocation) Assuming $L$ synced with $\mathbb{S}$: $\mathcal{A}$ is unable to provide 2 $\mathbb{C}$s with 2 distinct valid views on $L$. 
\end{theorem}
\begin{proof1}
	Since $L$ is regularly synced with publicly visible $\mathbb{S}$, and $\mathbb{S}$ stores only a single current version of $L$ (i.e., $LRoot_{pb}$), all $\mathbb{C}s$ share the same view on $L$.
\end{proof1}

\begin{theorem}\label{theorem:censorhip}
	(Censorship Evidence) $\mathcal{A}$ is unable to censor any request (tx/query) from $\mathbb{C}$ while staying unnoticeable.
\end{theorem}
\begin{proof1}
	If $\mathbb{C}$'s request is censored, $\mathbb{C}$ asks for a resolution of the request through public $\mathbb{S}$.
	$\mathcal{A}$ observing the request might either ignore it and leave the proof of censoring at $\mathbb{S}$ or submit the request to $\mathbb{E}$ and obtain an $\mathbb{E}$-signed proof witnessing that a request was processed -- this proof is submitted to $\mathbb{S}$, publicly resolving the request.
\end{proof1}

\vspace{-0.1cm}
\subsection{Other Properties and Implications}
\subsubsection{\textbf{Privacy VS Performance}}
\name provides privacy of data submitted to $\mathbb{S}$ during the censorship resolution since the requests and responses are encrypted.
However, it does not provide privacy against $\mathbb{O}$ who has the read access to $L$.
Although ledger could be designed with encryption, a disadvantage of such an approach would be the performance drop caused by the decryption of  requested data from $L$ upon every $\mathbb{C}$'s read query, requiring a call of $\mathbb{E}$.
In contrast, with partial-privacy, $\mathbb{O}$ is able to respond queries of $\mathbb{C}$s without touching $\mathbb{E}$.

\subsubsection{\textbf{Access Control at $\mathbb{S}$}}\label{sec:cens-write-access}
$\mathbb{C}$s interact with $\mathbb{S}$ only through functions for submission of censored requests.
Nevertheless, access to these functions must be regulated through an access control mechanism to avoid exhaustion (i.e., DoS) of this functionality by external entities. 
This can be performed with a simple access control requiring $\mathbb{C}$s to provide access tickets when calling the functions of $\mathbb{S}$.
An access ticket could be provisioned by $\mathbb{C}$ upon registration at $\mathbb{O}$, and it could contain $PK_{\mathbb{C}}^{pb}$ with a time expiration of the subscription, signed by $\mathbb{E}$.
Whenever $\mathbb{C}$ initiates a censored request, verification of an access ticket would be made by $\mathbb{S}$, due to which DoS of this functionality would not be possible.

\subsubsection{\textbf{Security of TEE}}
Previous research has shown that practical implementations of TEE, such as SGX can be vulnerable to memory corruption attacks~\cite{biondo2018guard} as well as side channel attacks~\cite{brasser2017dr,van2018foreshadow,Lipp2021Platypus,Murdock2019plundervolt}.
A number of software-based defense and mitigation techniques have been  proposed~\cite{shih2017t,gruss2017strong,chen2017detecting,brasser2017dr,seo2017sgx} and some vulnerabilities were patched by Intel at the hardware level~\cite{intel-sgx-response}.
Nevertheless, we note that \name is TEE-agnostic thus can be integrated with other TEEs (e.g., \cite{Keystone-enclave},~\cite{costan2016sanctum}).
Next, Matetic et al.~\cite{matetic2017rote} investigated rollback attacks on Intel SGX, enabling to repeatedly restart the state of an enclave.
The proposed protection technique ROTE provides state integrity by storing enclave-specific counters in a distributed system.
Note that \name deals with these attacks by regular snapshotting of the state of $\mathbb{E}$ to $\mathbb{S}$ (see \autoref{sec:failed-enclave}).
Another class of SGX vulnerabilities was presented by Cloosters et al.~\cite{cloosters2020teerex} and involved incorrect application designs enabling arbitrary reads and writes of protected memory or work done by Borrello et al. which involves serious microarchitectural flaws in chip design~\cite{Borrello2022AEPIC}. 
Since the authors have not published their tool, 
we did manual inspection of \name code and did not find any of the concerned vulnerabilities. 

\subsubsection{\textbf{Security vs. Performance}}
In SGX, performance is always traded for security, and our intention was to optimize the performance of \name while making custom security checks whenever possible instead of using expensive buffer allocation and copying to/from $\mathbb{E}$ by trusted runtime of SDK (\textit{trts}).
\textbf{Output Parameters:}
In detail, in the case of ECALL $Exec()$ function where $\mathbb{E}$ is provided with \texttt{[user\_check]} output buffers pointing to host memory, the strict location-checking is always made in $\mathbb{E}$ while assuming maximal size of the output buffer passed from the host (i.e., $oe\_is\_outside\_enclave(buf,~max)$).
Moreover, the maximal size is always checked before any write to such output buffers.
The concerned parameters of $Exec()$ are buffers for newly created and modified account state objects.
\textbf{Input Parameters:}
On the other hand, in the case of input parameters of $Exec()$, we utilize embedded buffering provided by trts of SDK since $\mathbb{E}$ has to check the integrity of input parameters before using them, otherwise Time-of-Check != Time-of-Use vulnerability~\cite{cloosters2020teerex} might be possible.
The concerned input parameters of $Exec()$ are transactions to process and their corresponding codes.

\subsubsection{\textbf{Time to Finality}}
Many blockchain platforms suffer from accidental forks, which temporarily create parallel inconsistent blockchain views.
To mitigate this phenomenon, it is recommended to wait a certain number of block confirmations after a given block is created before considering it irreversible with overwhelming probability.
This waiting time (a.k.a., time to finality) influences the non-equivocation property of \name, and it inherits it from the underlying blockchain platform.
Most blockchains have a long time to finality, e.g., $\sim$60 min in Bitcoin~\cite{nakamoto2008bitcoin}, $\sim$15mins in PoS Ethereum~\cite{wood2014ethereum} ($\sim$3 min in its former PoW version), $\sim$2min in Cardano~\cite{kiayias2017ouroboros}.
However, some blockchains (or consensus protocols) have a short time to finality, e.g., Avalanche~\cite{rocket2019scalable},
Algorand~\cite{gilad2017algorand},
Hedera~\cite{baird2019hedera},
Fantom~\cite{nguyen2021lachesis}, 
StrongChain~\cite{strongchain}. 
The selection of the underlying blockchain platform 
is dependent on the requirements of the particular use case.

\section{Implementation}
\label{sec:implementation}
We have made a proof-of-concept implementation of \name, where we utilized Intel SGX and C++ for instantiation of $\mathbb{E}$, while $\mathbb{S}$ was built on top of EVM-compatible blockchain. In detail, we utilized OpenEnclave SDK~\cite{open-enclave} and a minimalistic EVM, called eEVM~\cite{eEVM-Microsoft}.
However, eEVM is designed with the standard C++ map for storing the full state of $L$, which lacks efficient integrity-oriented operations.
Moreover, eEVM assumes the unlimited size of $\mathbb{E}$ for storing the full state, while the size of $\mathbb{E}$ in SGX is constrained to $\sim$100 MB.
This might work with enabled swapping but the performance of $\mathbb{E}$ is significantly deteriorated with a large full state.
Due to these limitations, we customized eEVM and replaced eEVM's full state handling with Merkle-Patricia Trie from Aleth~\cite{aleth}, which we modified to support operations with the partial state. 
$\mathbb{O}$ and $\mathbb{C}$ were also implemented in C++. 

Our implementation enables the creation and interaction of simple accounts as well as smart contracts written in Solidity. 
We verified the code of $\mathbb{S}$ by static/dynamic analysis tools Mythril~\cite{mythrill}, Slither~\cite{slither}, and ContractGuard~\cite{ContractGuard-fuzzer}. The source code of our implementation will be made available upon publication.

\begin{figure}[t]
	\centering	
	\vspace{-0.6cm}
	
	\subfloat[\label{fig:pay-TB} Turbo Boost enabled]{
		\hspace{-0.2cm}
		\includegraphics[width=0.23\textwidth]{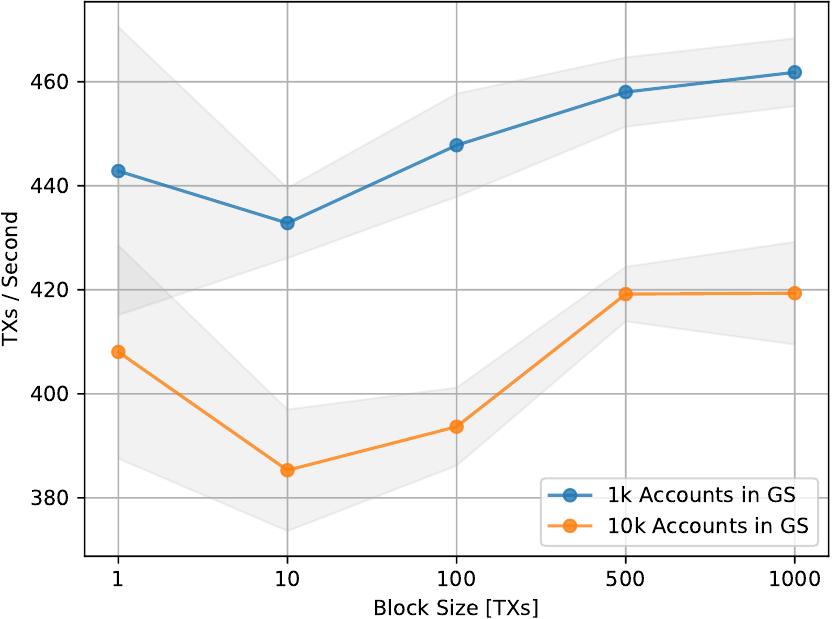} 
	}
	\subfloat[\label{fig:pay-noTB} Turbo Boost disabled]{
		\includegraphics[width=0.23\textwidth]{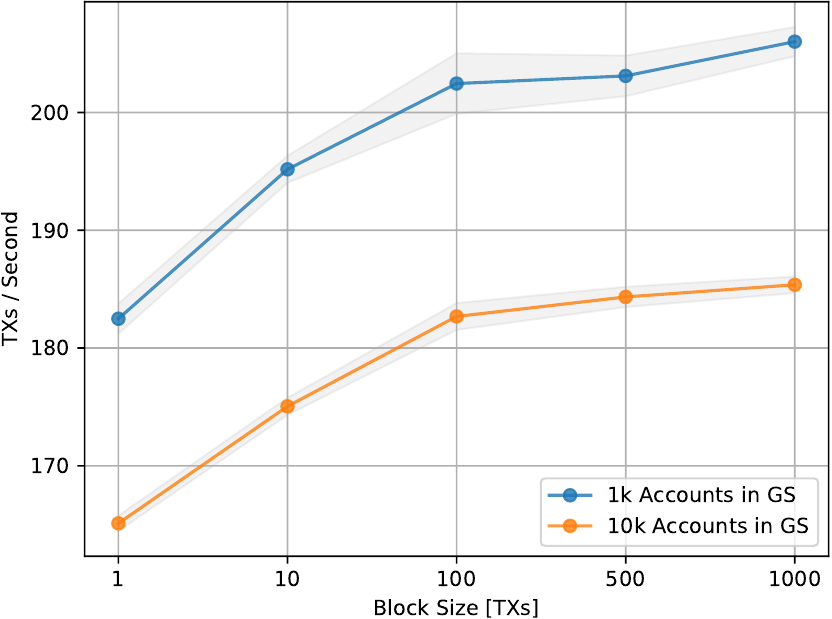} 
	}
	
	\vspace{-0.1cm}
	\caption{Performance of \name for native payments.}
	\label{fig:performance-enc-payments}
	\vspace{-0.3cm}
	
\end{figure}
\begin{figure}[t]
	\centering	
	\vspace{-0.2cm}
	
	\subfloat[\label{fig:erc-TB} Turbo Boost enabled]{
		\hspace{-0.2cm}
		\includegraphics[width=0.23\textwidth]{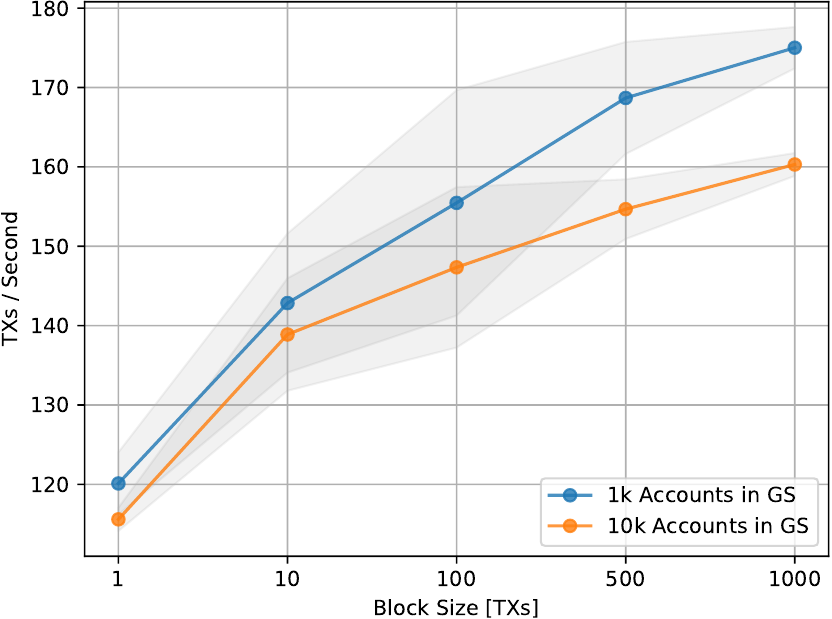} 
	}
	\subfloat[\label{fig:erc-noTB} Turbo Boost disabled]{
		\includegraphics[width=0.23\textwidth]{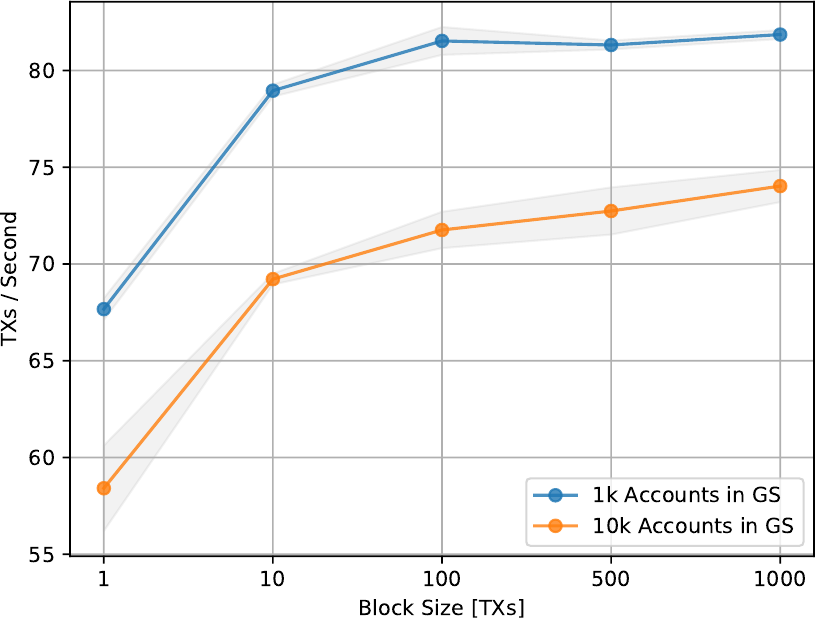} 
	}
	
\caption{Performance of \name for ERC20 smart contract calls.}
	\label{fig:performance-enc-erc}
	\vspace{-0.5cm}
\end{figure}

\subsection{Performance Evaluation}
All our experiments were performed on commodity laptop with Intel i7-10510U CPU supporting SGX v1, and they were aimed at reproducing realistic conditions -- i.e., they included all operations and verifications described in \autoref{sec:details}, such as verification of recoverable ECDSA signatures, aggregation of transactions by Merkle tree, integrity verification of partial state, etc.
We evaluated the performance of \name in terms of transaction throughput per second, where we distinguished transactions with native payments (see \autoref{fig:performance-enc-payments}) and transactions with ERC20 smart contract calls (see \autoref{fig:performance-enc-erc}).
All measurements were repeated 100 times, and we depict the mean and standard deviation in the graphs.

\subsubsection{\textbf{A Size of the Full State}}
The performance of \name is dependent on the size of data that is copied from $\mathbb{O}$ to $\mathbb{E}$ upon call of $Exec()$.
The most significant portion of the copied data is a partial state, which depends on the height of the MPT storing the full state. 
Therefore, we repeated our measurements with two different full states, one containing $1k$ accounts and another one containing $10k$ accounts.
In the case of native payments, the full state with 10k accounts caused a decrease of throughput by 7.8\%-12.1\% (with enabled TB) in contrast to the full state with 1k accounts.
In the case of smart contract calls, the performance deterioration was in the range 2.8\%-8.4\% (with enabled TB).

\begin{figure}[t]
	\centering	
	\vspace{-0.6cm}
	
	\subfloat[\label{fig:cens-tx-submit} Submit TX]{
		\hspace{-0.2cm}
		\includegraphics[width=0.23\textwidth]{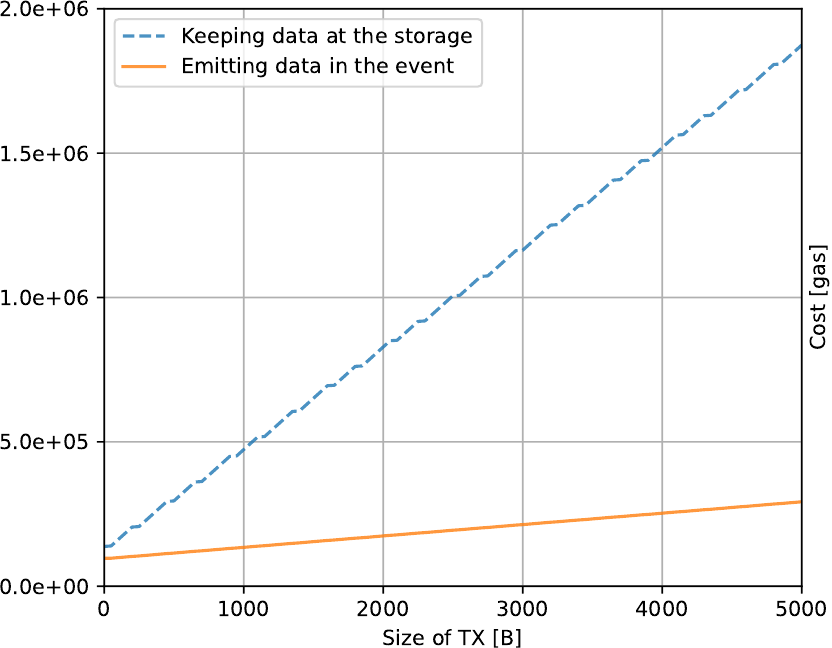} 
	}	
	\subfloat[\label{fig:cens-tx-resolve} Resolve TX]{
		\hspace{-0.0cm}
		\includegraphics[width=0.227\textwidth]{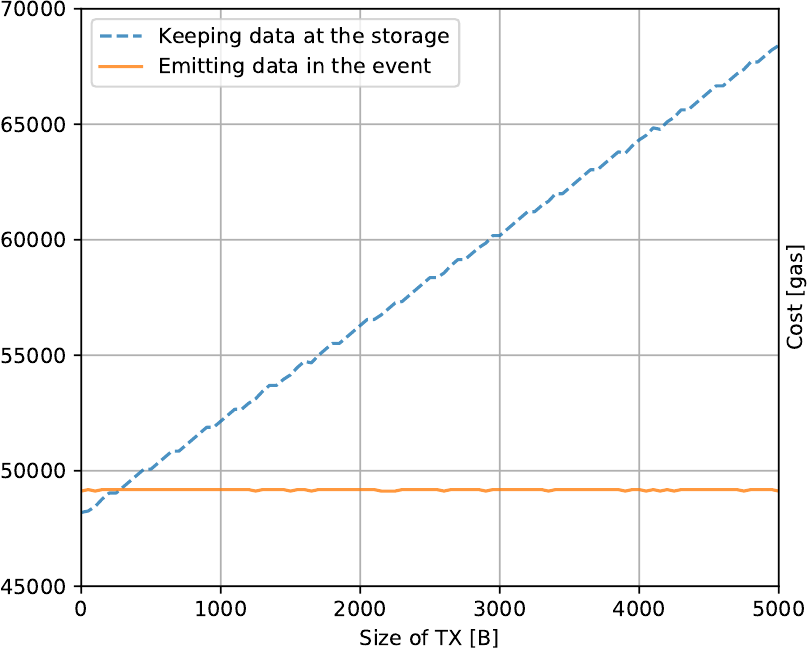} 
	}
	
	\vspace{-0.3cm}
	\subfloat[\label{fig:cens-qry-submit} Submit Query (Get TX)]{
		\includegraphics[width=0.23\textwidth]{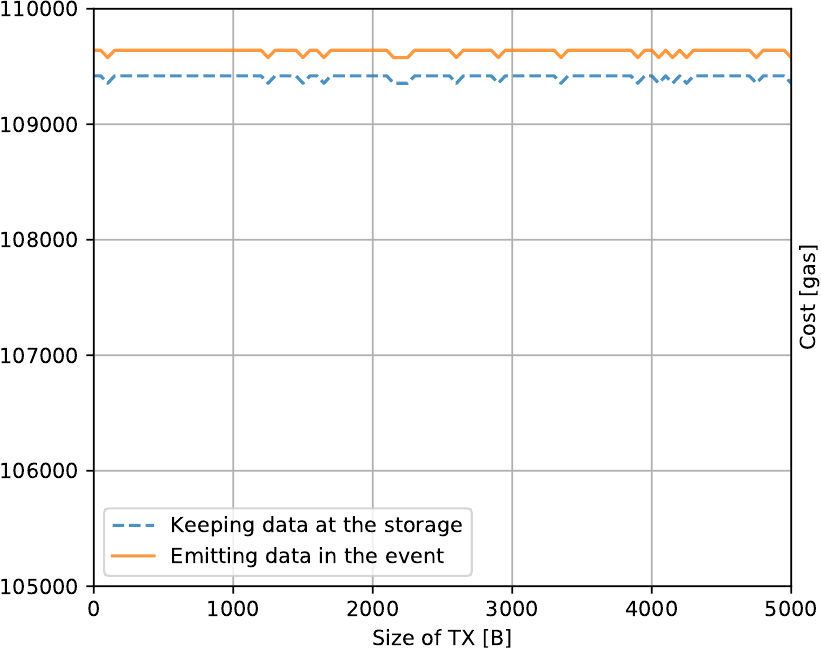} 
	}
	\subfloat[\label{fig:cens-qry-resolve} Resolve Query (Get TX)]{
		\includegraphics[width=0.23\textwidth]{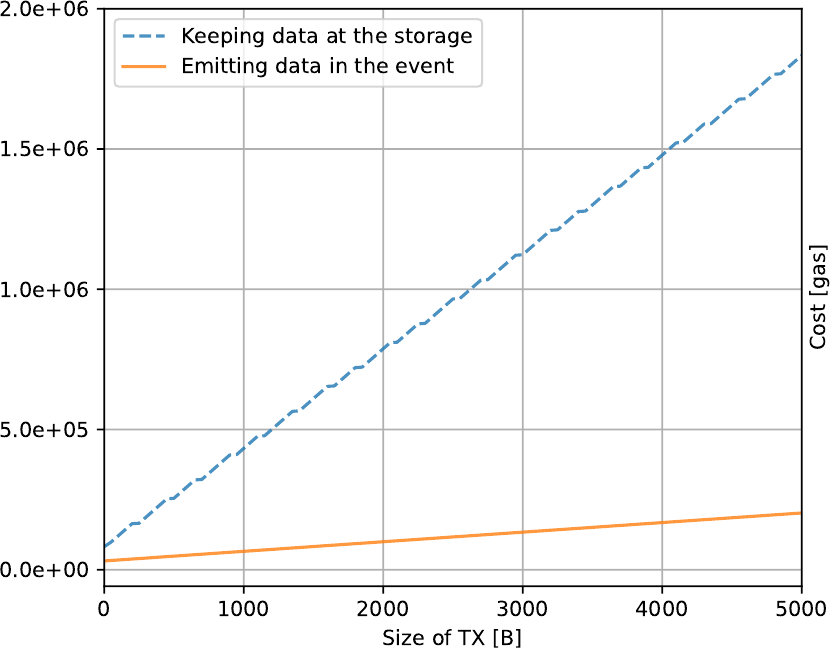} 
	}
	
\caption{Costs for resolution of censored txs and queries.}
	\label{fig:-submit-cens-queries}
	\vspace{-0.3cm}
\end{figure}

\subsubsection{\textbf{Block Size \& Turbo Boost}}
In each experiment, we varied the number of transactions aggregated in the block.
Initially, we performed measurements with enabled Turbo Boost (see \autoref{fig:pay-TB} and \autoref{fig:erc-TB}), where we witnessed a high throughput and its high variability. 
For smart contract calls (see \autoref{fig:erc-TB}), the throughput increased with the size of the block modified from 1 to 1000 by $45.7\%$ and $38.7\%$ for a full state with 1k and 10k accounts, respectively.  
However, in the case of native payments, the improvement was only $4.3\%$ and $2.8\%$, while the throughput was not increased monotonically with the block size.

Therefore, we experimentally disabled Turbo Boost (see \autoref{fig:pay-noTB}) and observed the monotonic increase of throughput with increased block size -- the improvement achieved was $11.41\%$ and $12.26\%$ for a full state with 1k and 10k accounts, respectively.
We also disabled Turbo Boost in the case of smart contract calls (see \autoref{fig:erc-noTB}), where the performance improvement was $20.9\%$ and $26.7\%$ for both full states under consideration.

\subsection{Analysis of Costs}\label{sec:cost-analysis}
Besides the operational cost resulting from running the centralized infrastructure,
\name imposes costs for interaction with the public blockchain with $\mathbb{S}$  deployed.
The deployment cost of $\mathbb{S}$ is $1.51M$ of gas and the cost of the most frequent operation -- syncing $L$ with $\mathbb{S}$ (i.e., $PostLRoot()$) -- is $33k$ of gas, which is only $33\%$ higher than the cost of a standard EVM-compatible transaction.
For example, if $L$ is synced with $\mathbb{S}$ every 5 minutes, $\mathbb{O}$'s monthly expenses for this operation would be $285M$ of gas, while in the case of syncing every minute, monthly expenses would be $1,425M$ of gas.

\subsubsection{\textbf{Censorship Resolution}}\label{sec:exps-censorship}
Our mechanism for censorship resolution imposes costs on $\mathbb{C}$s submitting requests as well as for $\mathbb{O}$ resolving these requests.
The cost of submitting a censored request is dependent on the size of the request/response and whether $\mathbb{S}$ keeps data of a request/response in the storage or whether it just emits an asynchronous event with the data (i.e., a cheaper option).
We measured the costs of both options and the results are depicted in \autoref{fig:-submit-cens-queries}.
Nevertheless, for practical usage, only the option with event emitting is feasible (see solid lines in \autoref{fig:-submit-cens-queries}).
\autoref{fig:cens-tx-submit} and \autoref{fig:cens-tx-resolve} depict the resolution of a censored transaction, which is more expensive for $\mathbb{C}$ than for $\mathbb{O}$, who resolves each transaction with constant cost $49$k of gas (see \autoref{fig:cens-tx-resolve}).
On the other hand, the resolution of censored queries is more expensive for $\mathbb{O}$ since she has to deliver a response with data to $\mathbb{S}$ (see \autoref{fig:cens-qry-resolve}), while $\mathbb{C}$ submits only a short query, e.g., get a transaction (see \autoref{fig:cens-qry-submit}).
This might potentially allow $\mathbb{C}$s to DoS $\mathbb{O}$.
Therefore, submission of a censored query by $\mathbb{C}$ needs to have a mechanism that avoid such a potential DoS, e.g., charging of $\mathbb{C}$ dependent on the expected size of the response passed as a parameter to $\mathbb{S}$.

\section{Related Work}
\label{sec:related}

\paragraph{\textbf{Append-Only Designs}}
The first line of research revolves
around authenticated append-only data structures.  Haber and
Stornetta~\cite{haber1990time} proposed a hash chain associated with
transactions, proving their order.  Subsequently, their work was
improved~\cite{bayer1993improving} by aggregating transactions in a Merkle tree,
allowing more efficient proofs and updates.  However, these constructions still
require $O(n)$ messages to prove that one version of the ledger is an extension
of another.  Crosby and Wallach~\cite{crosby2009efficient} introduced append-only
logs with $O(\log n)$-long incremental and membership proofs.  Certificate
Transparency (CT)~\cite{laurie2013certificate} deploys this data structure to
create a public append-only log of digital certificates supporting efficient
membership and extension proofs but with inefficient exclusion proofs. The idea of
CT's publicly verifiable logs was then extended to other applications, like
revocation transparency~\cite{laurie2012revocation}, binary
transparency~\cite{fahl2014hey}, or key transparency~\cite{melara2015coniks}.
The CT's base construction was further improved by systems combining an append-only Merkle tree with an ordered Merkle
tree~\cite{ryan2014enhanced,kim2013accountable} aiming to implement a variant of an authenticated append-only dictionary.  
Besides making all certificates visible and append-only, these constructions use a constructed key-value mapping to prove e.g., that a certificate is revoked, or that a given domain has a certain list of certificates.  
These systems provide more powerful properties than CT, but unfortunately, they have inefficient $O(n)$ proofs in verifying both properties of their logs at the same time (i.e., append-only ledger with the correct key-value mapping).
A construction of append-only dictionaries with succinct proofs was 
proposed in~\cite{tomescu2019transparency}.  
Despite achieving the desired properties, this construction relies on stronger cryptographic assumptions and requires a trusted setup. 
Moreover, the scheme has efficiency bottlenecks as proving time grows with the data, rendering it impractical. Finally, such schemes designed for key-value databases are unable to handle smart contracts.
LedgerDB~\cite{ledgerdb} is deployed on Alibaba Cloud and utilize time-stamping authority to regularly snapshot the ledger whose integrity is maintained using Merkle trees and hash chains.

\paragraph{\textbf{Non-Equivocation Designs}}
Although the above systems try to minimize trust in the operator of a ledger and aim at public verifiability by deploying cryptographic constructions, they require an out-of-band mechanism to provide non-equivocation.  
One family of solutions detecting equivocations are gossip protocols, where users exchange their ledger views to find any
inconsistencies~\cite{chuat2015efficient,dahlberg2018aggregationbased}.  
A disadvantage of these solutions is that they are primarily detective, unable to effectively prevent equivocation attacks, in contrast to \name.

Another approach to the non-equivocation of a ledger was
proposed in~\cite{kim2013accountable,basin2014arpki} that introduced multiple auditing nodes running a consensus protocol.
Mitigations of the equivocation also include systems built on top of a blockchain platform that provides non-equivocation by design. An advantage of those solutions is that they are as strong as the underlying blockchain platform and with some latency can prevent operator equivocations. 
Catena~\cite{tomescu2017catena} proposes a system where a
centralized log proves its non-equivocation by posting a sequence of integrity-preserving transactions in the Bitcoin blockchain for its updates.
However, it requires clients to obtain all Catena transactions and their number is linear with the number of log updates (in contrast, \name requires clients to obtain only the most recent state of $\mathbb{S}$).  
PDFS~\cite{guarnizo2019pdfs} reduces this overhead (to constant) by a smart contract that validates consistency with the past by incremental update of the ledger using the history tree data structure (similarly, as in \name); however, it does not guarantee the correct execution, which \name does as well as it makes snapshotting of the ledger much cheaper.
Custos~\cite{paccagnella2020custos} focuses on the detection of tampering with system logs and it utilizes the logger with TEE and decentralized auditors.
However, auditors must regularly perform audits to detect tampering, which is expensive and time-consuming. 
In contrast, \name provides instant efficient proofs of data genuineness upon request.

\paragraph{\textbf{Decentralized Designs with TEE}}
Several systems combine TEE with block\-chains, mostly to improve the lacking properties of blockchains like confidentiality
or throughput bottlenecks.  
Teechain~\cite{lind2017teechain} is a system where Bitcoin transactions can be executed off-chain in TEE enclaves.  
Another system is Ekiden~\cite{cheng2018ekiden}, which offloads smart contract execution to dedicated TEE-supported parties. 
These parties can execute smart contract transactions efficiently and privately and since they are agnostic to the blockchain consensus protocol the transaction throughput can be scaled horizontally.  
A similar approach is taken by Das et al.~\cite{das2019fastkitten} who propose FastKitten, an approach that enhances Bitcoin with Turing-complete smart contracts that are executed off-chain within the TEE of the operator.
In the similar line, Frassetto et al.~\cite{FrassettoJKKSFS23} propose POSE, the optimistic off-chain smart contract execution, while only a few operations are executed on-chain. POSE utilizes a pool of TEE-equipped operators, out of which only one is performing off-chain execution and the rest is snapshotting the state and optionally might replace the executing one in the case of its failure.
While POSE requires on-chain initialization and other operations for each smart contract (paid by clients), \name can process all smart contract operations with a ledger off-chain, producing an on-chain snapshot once in a while at the expense of the operator.

This category also involves decentralized privacy-preserving smart contract platforms such as Secret Network~\cite{secret-ppp}, Phala~\cite{phala-ppp}, Oasis~\cite{oasis-ppp}, Obsucuro~\cite{obscuro-ppp}, and Integritee~\cite{integritee-ppp}.  
These platforms execute smart contract code in a decentralized fashion using TEE at each node. Furthermore, some of these platforms might be vulnerable to non-TEE-related privacy and rollback attacks~\cite{jean2024sgxonerate}.
In sum, all the approaches from this category are decentralized designs with certain distribution redundancy that might impose high operational costs; this is orthogonal to a centralized design of \name.

\section{Conclusion}
\label{sec:conclusion}
In this paper, we proposed \name, a framework for centralized ledgers, which provides verifiability, non-equivocation, and censorship evidence.
To achieve these properties, we leveraged a unique combination of TEE Turing-complete smart contract platform.
We showed that \name is deployable with the current tools and can process over $400$ transactions per second on a commodity PC, including the overhead of all verifications.

\bibliographystyle{IEEEtran}
\bibliography{ref}

\appendix
\section{Appendix}
\label{sec:appendix}

\subsection{Data Availability}
We will release the source code of this work under open source license upon its publication.
Besides, this research does not contain any other data required to reproduce our results.

\subsection{\textbf{Radix and Merkle-Patricia Tries}}\label{appendix:mpt-background}
Radix trie serves as a key-value storage.
In the Radix trie, every node at the $l$-th layer of the trie has the form of $\langle (p_0, p_1, \ldots, p_n), v\rangle$, where $v$ is a stored value and all $p_i, ~i\in \{0,1, \ldots, n\}$ represent the pointers on the nodes in the next (lower) layer $l+1$ of the trie, which is selected by following the $(l+1)$-th item of the key.
Note that the key consists of an arbitrary number of items that belong to an alphabet with $n$ symbols (e.g., hex symbols). 
Hence, each node of the Radix trie has $n$ children and to access a leaf node (i.e., data $v$), one must descend the trie starting from the root node while following the items of the key one by one.
Note that Radix trie requires an underlying database of key-value storage that maps pointers to nodes.
However, the Radix trie does not contain integrity protection, and when its key is too long (e.g., hash value), the Radix trie will be sparse, thus imposing a high overhead for storage of all the nodes on the path from the root to values.

Merkle Patricia Trie (MPT)~\cite{wood2014ethereum,Merkle-Patricia-Trie-eth} is a combination of the Merkle tree (see \autoref{sec:MT-background}) and Radix trie data structures, and similar to the Radix Trie, it serves as a key-value data storage.
However, in contrast to the Radix trie, the pointers are replaced by a cryptographically secure hash of the data in nodes, providing integrity protection.
In detail, MPT guarantees integrity by using a cryptographically secure hash of the value for the MPT key as well as for the realization of keys in the underlying database that maps the hashes of nodes to their content; therefore, the hash of the root node of the MPT represents an integrity snapshot of the whole MPT trie.
Next, Merkle-Patricia trie introduces the \textit{extension nodes}, due to which, there is no need to keep a dedicated node for each item of the path in the key. 
The MPT trie $T$ supports the following operations:
\begin{compactdesc}
	\item[$\mathbf{T.root \rightarrow Root}$:] accessing the hash of the root node of MPT, which is stored as a key in the underlying database.
	
	\item[$\mathbf{T.add(k, x) \rightarrow Root}$:] adding the value $x$ with the key $k$ to $T$ while obtaining the new hash value of the root node. 
	
	\item[$\mathbf{T.get(k) \rightarrow \{x, \perp\}}$:] fetching a value $x$ that corresponds to key $k$; return $\perp$ if no such value exists.  
	
	\item[$\mathbf{T.delete(k) \rightarrow \{True, False\}}$:] deleting the entry with key equal to $k$, returning $True$ upon success, $False$ otherwise.
	
	\item[$\mathbf{T.MptProof(k) \rightarrow  \{\pi^{mpt}, \pi^{\overline{mpt}}\}}$:] an MPT (inclusion/exclusion) proof generation for the entry with key $k$.

	\item[$\mathbf{\pi^{mpt}.Verify(k, Root) \rightarrow \{True, False\}}$:]  verification of the MPT proof $\pi^{mpt}$, witnessing that entry with the key $k$ is in the MPT whose hash of the root node is equal to $Root$.			
	
	\item[$\mathbf{\pi^{\overline{mpt}}.VerifyNeg(k, Root) \rightarrow \{True, False\}}$:]  verification of the negative MPT proof, witnessing that entry with the key $k$ is not in the MPT with the root hash equal to $Root$.			
\end{compactdesc}

\subsection{\textbf{Examples of Censored Queries}}\label{sec:censored-query-example}
While in \autoref{sec:censored-query} and \autoref{fig:censored-query} we omit the details about the data that a query might fetch, here we describe two examples.

\paragraph{\textbf{Get Transaction}}
In the first example, a query fetches the transaction $tx$ identified by $id_{tx}$ that is part of the block identified by $id_{blk}$.\footnote{To verify whether the block with $id_{blk}$ exists, we check $id_{blk} \leq$  $\#(LRoot_{pb})$.}
Upon notification from $\mathbb{S}$ about the unresolved request,  $\mathbb{O}$ fetches the full block with ID equal to $id_{blk}$,\footnote{Note that a full block is required to pass into $\mathbb{E}$ since Merkle tree (aggregating transactions) does not support exclusion proofs, and thus all transactions of the block need to be compared.} 
computes its membership proof $\pi_{hdr}^{mem}$ in the version $\#(LRoot_{pb})$ of $L$, and calls the function $SignQryTx()$ of $\mathbb{E}$ (see \autoref{alg:enclave-VM-cens}) with these data in the arguments.
$\mathbb{E}$ verifies $\pi_{hdr}^{mem}$ and searches for $tx$ with $id_{tx}$ in the passed block.
If $tx$ is found, $\mathbb{E}$ signs encrypted $tx$ and the positive status of the query.
On the other hand, if $tx$ is not found in the block, $\mathbb{E}$ signs the negative query status and empty data.
The signature, the status, and encrypted $tx$ are passed to $\mathbb{S}$, where the censorship of the query is finished.

\paragraph{\textbf{Get Account State}}
In the second example, a query fetches an account state $as$ identified by $id_{as}$ from the most recent version $\#(LRoot_{cur})$ of $L$. 
When $\mathbb{O}$ is notified by $\mathbb{S}$ about an unresolved request, $\mathbb{O}$ retrieves $as$ from MPT trie storing the full global state of $L$, computes its MPT proof $\pi^{mpt}_{as}$,\footnote{If $as$ is not found, $\pi^{mpt}_{as}$ serves as a negative proof of $as$.} and calls the function $SignQryAS()$ of $\mathbb{E}$ (see \autoref{alg:enclave-VM-cens}) with these data in the arguments.
$\mathbb{E}$ verifies $\pi^{mpt}_{as}$ with regards to $\#(LRoot_{cur})$, and if it is a positive MPT proof, $\mathbb{E}$ signs the encrypted $as$ and a positive status of the query.
In contrast, if $\pi^{mpt}_{as}$ is a negative MPT proof, $\mathbb{E}$ signs the negative query status and the empty data.
The signature, status, and encrypted $as$ are passed to $\mathbb{S}$, where the censorship of the query is completed.

\subsection{\textbf{Universally Composable Security}}\label{appendix:uc-proof}
The goal is to prove that the $\proto_{AQ}$ realizes $\AQledger$ assuming the 
EUF-CMA security of the underlying signature scheme, a collision-resistant hash function and security properties of the used proofs, \ie Merkle, incremental, and membership. 
We construct a simulator $\IAdv$ that simulates the real execution of $\proto_{AQ}$ to $\adv$. The view of the real execution, provided by $\IAdv$ to $\adv$  cannot be distinguishable to a PPT $\env$  from the ideal execution, \ie execution between $\AQledger$ and $\IAdv$. First, we outline the interface of the protocol.

\subsubsection{\textbf{Interface of $\mathbb{C}$ and $\mathbb{O}$}}
To model $\proto_{AQ}$ within the UC framework, we rewrite the interface of the $\mathbb{C}$ and $\mathbb{O}$ as follows: 
\begin{compactdesc}
	\item \textbf{Operator $\op$}
	\begin{compactitem}
		\item On the input of $\msg{Init}{}$,\footnote{Note that $sid$ represents the session ID of the instance of $\proto_{AQ}$.} it performs the setup procedures $\proto_{S}$ via \textit{write} interaction with the public blockchain to store the PKs of $\mathbb{E}$ and $\mathbb{O}$; It outputs $\msg{initOK}{}$; 

		\item On the input of $\msg{Publish}{}$, it publishes the transactions within a batch (\ie $block$ of $\ledger$) via the \textit{write} interaction with the public blockchain to snapshot the version of $\ledgerStateList$ containing executed transactions and then outputs $\msg{PublishOK}{}$.
\end{compactitem}	
	
	\item \textbf{Client $\client_i$}
	\begin{compactitem}
		\item On the input of $\msg{Add}{\tx}$ for a transaction $\tx$, it performs the protocol $\proto_N$. It also involves the \textit{read} interaction with the public blockchain in terms of checking the snapshotted/irreversible version of $\ledgerStateList$ and outputs $\msg{AddOk}{\tx}$; 

		\item On the input of $\msg{Receipt}{\tx}$ for a transaction $\tx$, it performs the protocol $\proto_R$, involving the \textit{read} interaction with the public blockchain to obtain the recent version of $\ledgerStateList$. Then it outputs the message $\msg{ReceiptOk}{\phi}$ for $\phi=\{\mbox{``Receipt valid''},$ $\mbox{``Receipt invalid''}\}$;\footnote{The invalid receipt represents not processed (\ie censored) transaction.}

		\item On the input of $\msg{ResolveQuery}{\qry}$ for a query $\qry$, it performs $\proto_{CQ}$, which involves the \textit{read} interaction with the public blockchain, and outputs $\msg{ResolveQueryOK}{\qry(\ledgerStateList),\mathsf{msg}}$ such that $\mathsf{msg}$ is the auxiliary message
from the TEE, \eg an error message from TEE, and  $\qry(\ledgerStateList)$ is the result of the query $\qry$ over the ledger $\ledgerStateList$;
		
		\item On the input of $\msg{Read}{}$, it verifies the public blockchain for processed transactions. Then it outputs $\msg{ReadOk}{\ledgerStateListPub}$ with the most recent state of the internal ledger snapshotted by the public blockchain;
		
		\item On the input of $\msg{ResolvedQueries}{}$, it verifies the public blockchain w.r.t. the resolved queries. 
		Then it outputs the message with the state $\mathsf{result}$ in the public blockchain, $\msg{ResolvedQueriesOK}{\mathsf{result}}$; 

		\item On the input of $\msg{HandledReceipts}{}$, it verifies the public blockchain for receipts. Then  it outputs $\msg{HandledReceiptsOK}{\mathsf{result}}$ with the state of the public blockchain;
		
		\item On the input of $\msg{Query}{\qry}$, the operator performs the query $\qry$ on the kept ledger. Then it outputs $\msg{QueryOk}{\mathsf{result}}$ with the result of the query.
		
	\end{compactitem}
\end{compactdesc}

\setcounter{theorem}{0}
\begin{theorem}
Assuming a collision-resistant hash function, security of Merkle, incremental and membership proofs, and the EUF-CMA signature scheme for the enclave $\mathbb{E}$, them $\proto_{AQ}$ realizes $\AQledger$ in the presence of adversary $\adv$ who corrupts the operator $\op$ but not the enclave $\mathbb{E}$. 
\end{theorem}    

\begin{proof}
	The proof is based on the construction of a simulator $\IAdv$ which interacts with $\AQledger$ in the ideal execution. We later show that the generic construction of $\IAdv$ renders the ideal execution between $\IAdv$ and $\AQledger$ as well as the real execution between $\adv$ and  $\proto_{AQ}$ computationally indistinguishable for a PPT $\env$, given an EUF-CMA signature scheme which generates the enclave key pair $(\mathsf{pk}^{pb}_{\mathbb{E}},\mathsf{sk}^{pb}_{\mathbb{E}})$.  We start by providing the construction for $\IAdv$.

	\paragraph{Construction of $\IAdv$} The simulator  $\IAdv$ uses  adversary $\adv$ attacking the protocol  $\proto_{AQ}$. Concretely,  $\IAdv$ generates $(\mathsf{pk}^{pb}_{\mathbb{E}},\mathsf{sk}^{pb}_{\mathbb{E}})$, $(\mathsf{pk}^{TEE}_{\mathbb{E}},\mathsf{sk}^{TEE}_{\mathbb{E}})$ and $(\mathsf{pk}^{pb}_{\op},\mathsf{sk}^{pb}_{\op})$ via an EUF-CMA signature scheme. The key pairs are to be used by $\IAdv$'s simulation of the enclave, the public blockchain, internally, and by $\op$, respectively. Furthermore, $\IAdv$ provides the simulation of  $\proto_{AQ}$, by offering a view of $n$ simulated clients $\client_{i}$ interacting with a simulated $\op$, and a simulated public blockchain. The adversary $\adv$ is provided with the exchange of messages between the protocol participants (by $\IAdv$). When $\adv$ sends a message to corrupt $\op$, $\IAdv$ provides the internal private  information of the simulated $\op$ and let $\adv$ output actions for the corrupted $\mathbb{O}$ as follows:
	
	\begin{compactitem}
		\item Upon receiving $\msg{corrupt}{}$ from $\adv$, submit $\msg{corrupt}{}$ to $\AQledger$ and hand the private information of the simulated $\mathbb{O}$ to $\adv$, \ie $\mathsf{sk}^{pb}_{\op}$. It is important to highlight that $\IAdv$ does not hand the private information with respect to TEEs, \ie  $\mathsf{sk}^{pb}_{\mathbb{E}}$ and $\mathsf{sk}^{TEE}_{\mathbb{E}}$, to~$\adv$.
	\end{compactitem}	
	\ \\
	In the following, we thoroughly describe $\IAdv$'s actions.
	\begin{compactitem}
		\item Upon receiving $\msg{Add}{\tx,\client_i}$ from $\AQledger$, simulate the normal traffic of messages via the simulated protocol to $\adv$. Furthermore, if $\mathbb{O}$ is corrupted, then $\adv$ controls its actions. Therefore, input $(Add,\tx)$ to $\adv$ which outputs $\action\in\{0,1\}$ and $\mathsf{msg}^\prime$. Thus, send $\msg{Add}{\action,\mathsf{msg}^\prime}$ to $\AQledger$;
		
		\item Upon receiving $\msg{Receipt}{\tx,\client_i}$ from $\AQledger$,  $\IAdv$ simulates the view of the public blockchain and the message exchange to  $\adv$. If $\tx$ had been previously received and published in the internal ledger, then return $\msg{Receipt}{\action=1}$, otherwise return $\msg{Receipt}{\action=0}$. In the case the Operator was previously corrupted, $\adv$ controls its actions. Therefore, input $(Receipt,\tx)$ to $\adv$ which outputs $\action\in\{0,1\}$, while $\adv$ generates the signature on the receipt and returns it to the simulated public blockchain by $\IAdv$.  Then return  $\msg{Receipt}{\action}$ to  $\AQledger$;

		\item Upon receiving $\msg{Query}{\qry,\client_i}$ from $\AQledger$, input $(\client_i,\qry)$ to $\adv$. When $\adv$ outputs $\mathsf{result}^\prime$, then return   $\msg{Query}{\mathsf{result}^\prime,\client_i}$ to $\AQledger$;

		\item  If $\op$ is honest, upon receiving $\msg{Publish}{\batch}$ from $\AQledger$, simulate the view of the public blockchain and the message exchange to  $\adv$. 
		If $\op$  is corrupted, then submit $\batch$ to $\adv$ and wait until it outputs $\batch^\prime$. Then, simulate public blockchain with $\batch^\prime$ and reply $\msg{Publish}{\batch^\prime}$ to $\AQledger$;

		\item If $\op$ is corrupted, upon receiving $\msg{ResolveQuery}{\qry,\client_i}$, input  $(\client_i,\qry)$ to $\adv$ and let it output $(\action,\mathsf{result}^\prime, \mathsf{msg}^\prime)$. Then, return to $\client_i$ the message $\msg{ResolveQuery}{\qry,\mathsf{result}^\prime, \mathsf{msg}^\prime}$.

	\end{compactitem}

	\paragraph{Security analysis of $\IAdv$} Given the earlier construction of simulator $\IAdv$, note that it perfectly simulates the operation of $\proto_{AQ}$ in the case $\op$ is not corrupted. 
In the case $\mathbb{O}$ is corrupted, consider the following cases where $\env$ drives the execution of the protocol:
	\begin{compactitem}
		\item \textbf{Censored Transaction/Valid Receipt:} Upon receiving $\msg{Add}{\tx^\ast,\client_i}$, for an arbitrary transaction  $\tx^\ast$, $\adv$ answers with $\action=0$ (\ie censored transaction). Upon the message   $\msg{Receipt}{\tx^\ast,\client_i}$, $\adv$ answers with $\action=1$ (\ie process receipt). In the case $\adv$  generates a valid receipt for the simulation of the protocol, let this event be denoted $\mathsf{Event\mbox{-}receipt}$; 
		
		\item \textbf{Successfully Tampered Batch:} Upon receiving $\msg{Publish}{\batch}$, $\adv$ outputs $\batch^\ast$. If   $\batch\neq\batch^\ast$, that means at least one $\tx^\ast$ differs. Upon the message $\msg{Receipt}{\tx^\ast,\client_i}$, $\adv$ answers with $\action=1$ (\ie process receipt); in the case $\adv$ generates a valid receipt for the simulation of the protocol, denote this event $\mathsf{Event\mbox{-}batch}$;
		1		
		
		\item \textbf{Resolved Tampered Query:}
Without loss of generality assume that $\IAdv$ keeps an honest ledger $\ledgerStateList^\IAdv$ for its simulation. Upon receiving the message  $\msg{Query}{\qry,\client_i}$,  and input $(\client_i,\qry)$ by $\adv$,  $\adv$ outputs $\mathsf{result}^\ast$. 
		If   $\qry(\ledgerStateList^\IAdv)\neq\mathsf{result}^\ast$, and $\mathsf{result}^\ast$ is accepted by the public blockchain simulation by $\IAdv$, then let this event be denoted $\mathsf{Event\mbox{-}query}$.
	\end{compactitem}

		\begin{algorithm}[t] 
		\caption{\footnotesize Censorship resolution by $\mathbb{O}$ (part of $prog^{\mathbb{O}}$)}\label{alg:operator-cens}
		\scriptsize 
		
		\SetKwProg{func}{function}{}{}
		
		\func{$UponPostedCensTX$($etx, idx_{req}$) }{
			$tx \leftarrow prog^{\mathbb{E}}.Decrypt(etx)$; \\
			$censTxs.add(\{tx, etx, idx_{req}\})$; \\
			$UponRecvTx$($tx$); \Comment{Delay response until the current block is finished.} \\
		}
		\smallskip		
		
		\func{$UponPostedCensQry$($equery, idx_{req}$) }{
			$qry \leftarrow parse(prog^{\mathbb{E}}.Decrypt(equery))$; \\
			\If{READ\_TX = $qry.type$}{
				
				$blk \leftarrow getBlockById(qry.id_{blk})$; \\
				$\pi^{mem}_{hdr} \leftarrow L.MemProof(blk.hdr.ID, LRoot_{pb}) $; \\
				
				$\sigma, status, edata \leftarrow prog^{\mathbb{E}}.SignQryTx(equery, blk,~ \pi^{mem}_{hdr})$; \\
			}\ElseIf{READ\_AS = $qry.type$}{
				$as \leftarrow state_{cur}.get(qry.id_{as})$; \\			
				$\pi^{mpt}_{as} \leftarrow L.MptProof(qry.id_{as})$; \Comment{Inclusion/exclusion proof.} \\		 		 		
				$\sigma, status, edata \leftarrow prog^{\mathbb{E}}.SignQryAS(equery, as, \pi^{mpt}_{as})$; \\
			}
			
			$prog^{\mathbb{S}}.ResolveCensQry(idx_{req}, status, edata, ~\sigma)$; \\
		}
		\smallskip			
		
		\func{$resolveCensTxs$() }{		
			\For{$\{ct:~  censTxs\}$}{										
				$blk \leftarrow getBlockOfTx(h(ct.tx), L)$; \\			
				$\pi^{mem}_{hdr} \leftarrow L.MemProof(blk.hdr.ID, LRoot_{pb}) $; \\
				$\pi^{mk}_{tx} \leftarrow MkProof(ct.tx, blk.txs) $; \\		
				
				$\sigma, status \leftarrow prog^{\mathbb{E}}.SignTx(ct.etx,~ \pi^{mk}_{tx},~ blk.hdr,~ \pi^{mem}_{hdr})$; \\
				
				$prog^{\mathbb{S}}.ResolveCensTx(ct.idx_{req}, status, ~\sigma)$; \\
			}				
			$censTxs \leftarrow []$; \\	
		}
		
	\end{algorithm}

	\paragraph{Security event probabilities}
	From now we estimate the probability of each of the early described events:
	\begin{compactitem}
		\item \textbf{Probability of $\mathsf{Event\mbox{-}receipt}$}: In such event, it means that $\adv$ could issue a valid signature $\sigma_{last}$ with enclave's private key to produce a valid receipt. This event produces two different outcomes between the real (the one with $\proto_{AQ}$) and ideal (the one with $\AQledger$) executions, and therefore $\env$ can distinguish between the two executions.  However  recall that the  $\mathsf{sk}^{pb}_{\mathbb{E}}$ was not disclosed to $\adv$, which means that  $\adv$ could 
generate a forgery, \ie $\sigma_{last}$, for the EUF-CMA signature scheme. Moreover, the receipt forgery, as given by protocol $\proto_{R}$, involves either one of the following cases:
		\begin{compactitem}
			\item forgery  of the incremental proof $\pi^{inc}$, \ie history tree;
			\item forgery  of membership proof $\pi^{men}_{hdr}$;
			\item forgery of Merkle tree proof $\pi^{mk}_{rcp_{i}}$;
			\item and a finding a collision for the hash of the transaction,
		\end{compactitem}
		in addition to the mentioned forgery of the signature scheme. Given the assumptions, \ie security of the proofs, a collision resistance hash function, and the  EUF-CMA security of the pair $(\mathsf{pk}^{pb}_{\mathbb{E}},\mathsf{sk}^{pb}_{\mathbb{E}})$, thereby we conclude this event happens with negligible probability;
		
		\item  \textbf{Probability of $\mathsf{Event\mbox{-}batch}$}: Alike the previous case, such event also requires $\adv$ to generate a forgery signature for $\mathsf{pk}^{pb}_{\mathbb{E}}$  to produce a valid receipts for  non executed tran\-sactions. 
		Therefore, it also happens with negligible probability given the EUF-CMA security of the signature scheme;
		
		\item  \textbf{Probability of  $\mathsf{Event\mbox{-}query}$}: If a tampered query was accepted by the simulated public blockchain (kept by the simulator $\IAdv$), then it means that once again $\adv$ somehow could generate a forgery for $\mathsf{pk}^{pb}_{\mathbb{E}}$ as described by protocol $\proto_{CQ}$. Thus, we conclude, similarly to the previous cases, that the probability of this event is negligible given the EUF-CMA security.	
	\end{compactitem}
	
	\ \\
	Given the earlier constructed simulator $\IAdv$ and its success probability, we conclude $\proto_{AQ}$ realizes $\AQledger$.	
\end{proof}

\subsection{\textbf{Additional Pseudo Codes of $\mathbb{O}$ and $\mathbb{E}$}}\label{appendix:pseudo-codes}
Although our protocols were principally described in the main text, here we provide additional pseudo code of $\mathbb{O}$'s normal operation in \autoref{alg:operator} and handling of censorship requests in \autoref{alg:operator-cens}.
Next, the pseudo code of censorship resolution by $\mathbb{E}$ is shown in \autoref{alg:enclave-VM-cens}.

\begin{algorithm}[t] 
	\scriptsize
	\SetKwProg{func}{function}{}{}

	\func{$Decrypt(edata)$ \textbf{public}}{
		$data \leftarrow \Sigma_{pb}.Decrypt(SK_\mathbb{E}^{pb}, edata)$; \\
		
		\textbf{Output}($data$); \\
	}
	\smallskip
	
	\func{$SignTx(etx, \pi^{mk}_{tx}, hdr, \pi^{mem}_{hdr})$ \textbf{public}}{
		\Comment{Resolution of a censored write tx. \hfill \hfill \hfill \hfill} \\
		
		$tx \leftarrow \Sigma_{pb}.Decrypt(SK_\mathbb{E}^{pb}, etx)$; \\
		\If{$ERROR = parse(tx)$}{					
			$status$ = PARSING\_ERROR;\\
			
		} \ElseIf{$ERROR = \Sigma_{pb}.Verify((tx.\sigma, tx.PK_{\mathbb{C}}^{pb}), tx) $}{
			$status$ = SIGNATURE\_ERROR;\\	
			
		} \Else{		
			\Comment{Verify proofs binding TX to header and header to $L$.\hfill \hfill}\\
			\textbf{assert} $\pi_{tx}^{mk}$.Verify($tx, hdr.txsRoot$);\\
			
			\textbf{assert} $\pi_{hdr}^{mem}$.Verify($hdr.ID, hdr, LRoot_{pb}$);\\					
			$status \leftarrow$  INCLUDED;\\
		}

		\Comment{TX was processed, so $\mathbb{E}$ can issue a proof.}\hfill \hfill \hfill \\		
		$\sigma \leftarrow \Sigma_{pb}.sign(SK_{\mathbb{E}}^{pb},~ (h(etx), status))$;  \\
		\textbf{Output}($\sigma$, $status$); 				
	}		
	\smallskip	
	
	\func{$SignQryTx(equery, blk, \pi^{mem}_{hdr})$ \textbf{public}}{
		\Comment{Resolution of a censored read tx query. \hfill \hfill \hfill \hfill} \\
		
		$\ldots, id_{tx}, id_{blk}, PK_{\mathbb{C}}^{pb} \leftarrow parse(Decrypt(equery))$; \\		
		\If{$id_{blk} > \#(LRoot_{pb})$}{
			$status \leftarrow$  BLK\_NOT\_FOUND, $~~edata \leftarrow$  $\perp$;\\
		}\Else{  
			
			\textbf{assert} $\pi_{hdr}^{mem}$.Verify($blk.hdr.ID, blk.hdr, LRoot_{pb}$);\\
			
			\textbf{assert} VerifyBlock(blk); \Comment{Full check of block consistency.} \\				
			
			$tx \leftarrow$ findTx$(id_{tx}, blk.txs)$; \\
			\If{$\perp ~=~ tx$}{
				$status \leftarrow$  TX\_NOT\_FOUND, $~~edata \leftarrow$  $\perp$;\\
			}\Else{
				$status \leftarrow$  OK, $~~edata \leftarrow  \Sigma_{pb}.Encrypt(PK_\mathbb{C}^{pb}, tx)$;\\
			}
		}
		
		$\sigma \leftarrow \Sigma_{pb}.sign(SK_{\mathbb{E}}^{pb},~ (h(equery), status, edata))$;  \\	
		\textbf{Output}($\sigma$, $status$, $edata$); 	
	}		
	
	\func{$SignQryAS(equery, as, \pi^{mpt}_{as})$ \textbf{public}}{
		\Comment{Resolution of a censored read account state query. \hfill \hfill \hfill \hfill} \\
		
		$\ldots, id_{as}, PK_{\mathbb{C}}^{pb} \leftarrow parse(Decrypt(equery))$; \\				
		\If{$\perp ~=~ as$}{
			\textbf{assert} $\pi^{mpt}_{as}.VerifyNeg(id_{as}, LRoot_{cur})$; \\
			$status \leftarrow$  NOT\_FOUND, $~~edata \leftarrow$  $\perp$;\\
		}\Else{
			\textbf{assert} $\pi^{mpt}_{as}.Verify(id_{as}, LRoot_{cur})$; \\
			$status \leftarrow$  OK, $~~edata \leftarrow  \Sigma_{pb}.Encrypt(PK_\mathbb{C}^{pb}, as)$;\\
		}
		
		$\sigma \leftarrow \Sigma_{pb}.sign(SK_{\mathbb{E}}^{pb},~ (h(equery), status, h(edata)))$;  \\	
		\textbf{Output}($\sigma$, $status$, $edata$); 	
	}		
	
	\caption{\footnotesize Censorship resolution in $\mathbb{E}$ (part of $prog^{\mathbb{E}})$.}\label{alg:enclave-VM-cens}
	
\end{algorithm}

\begin{algorithm}[ht] 
	\caption{\footnotesize The program $prog^{\mathbb{O}}$ of operator $\mathbb{O}$}\label{alg:operator}
	\scriptsize
	
	\SetKwProg{func}{function}{}{}
	
	$\triangleright$ \textsc{Variables and functions of $\mathbb{O}$:} \\	
	\hspace{1em} $PK_{\mathbb{E}}^{tee}, PK_{\mathbb{E}}^{pb}$: public keys of enclave $\mathbb{E}$ (under $\Sigma_{tee}$ $\&$ $\Sigma_{pb}$),\\
	\hspace{1em} $PK_{\mathbb{O}}, SK_{\mathbb{O}}$: keypair of operator $\mathbb{O}$ (under $\Sigma_{pb}$),\\
	\hspace{1em} $prog^{\mathbb{E}}, prog^{\mathbb{S}}$: program of enclave/smart contract, \\
	\hspace{1em} $txs_{u}$: cache of unprocessed TXs, \\
	\hspace{1em} $blks_{p}^{\#}$: counter of processed blocks, not synced with PB yet,\\			
	\hspace{1em} $\tau_{vm}, \tau_{pb}$: time of the last flush to  enclave/PB,\\			
	\hspace{1em} $state_{cur} \leftarrow \perp $: current full global state of VM,\\		
	\hspace{1em} $censTxs \leftarrow [] $: cache of posted censored TXs to $\mathbb{S}$,\\					
	\hspace{1em} $L \leftarrow []$: data of $L$ (not synced with PB),\\							
	\hspace{1em} $LRoot_{pb}$: the last root of $L$ synced with PB,\\
	\hspace{1em} $LRoot_{cur}$: the current root of $L$ (not synced with PB),\\
	\hspace{1em} $\sigma_{last}$: a signature of the last version transition pair signed by $\mathbb{E}$,\\

	\smallskip
	$\triangleright$ \textsc{Declaration of types and constants:}\\		
	\hspace{1em} \textbf{Block} \{$hdr, txs, rcps$\};  \\
	\hspace{1em} $FL^{\#}_{vm}, FL^{\#}_{pb}$: $\#$ of txs/blocks for flushing to enclave/PB, \\
	\hspace{1em} $FL^{\tau}_{vm}, FL^{\tau}_{pb}$: timeout for flushing to enclave/PB, \\
	\smallskip
	$\triangleright$ \textsc{Declaration of functions:}
	
	\func{$Init$() }{
		$PK_{\mathbb{E}}^{tee}, PK_{\mathbb{E}}^{pb} \leftarrow  prog^{\mathbb{E}}.Init()$;\\
		
		$prog^{\mathbb{S}}.Init(PK_{\mathbb{E}}^{pb}, PK_{\mathbb{E}}^{tee}, PK_{\mathbb{O}})$; \\		
		
	}
	\smallskip

	\func{$UponRecvTx$($tx$)  }{
		\textbf{assert} accessControl($tx$); \\
		$txs_{u}$.add($tx$); \\
		
		\If{$|txs_{u}| =  FL^{\#}_{vm} ~\vee~ now() - FL^{\tau}_{vm} \geq \tau_{vm}$}{
			
			$\pi^{inc}_{next}, LRoot_{tmp} \leftarrow nextIncProof()$;\\
			
			$LRoot_{pb}, LRoot_{cur}, \partial st^{new}, hdr, rcps, txs_{er}, \sigma_{last} \leftarrow$ $prog^{\mathbb{E}}.Exec(txs_{u}, \partial state^{cur}, ~\pi^{inc}_{next}, LRoot_{tmp})$; \\
			
			$state_{cur}.update(\partial st_{new})$;\\
			$L$.add(\textbf{Block($hdr$, $txs_{u} \setminus txs_{er}, rcps$)});\\
			$txs_{u} \leftarrow []$;				
			$~blks_{p}^{\#} \leftarrow blks_{p}^{\#} + 1$ ; \\

			\smallskip
			\Comment{Sync with $\mathbb{S}$ on public blockchain \hfill \hfill} \\
			\If{$blks_{p}^{\#} = FL^{\#}_{pb} ~\vee~ now() - FL^{\tau}_{pb} \geq \tau_{pb}$}{								
				
				$prog^{\mathbb{S}}.PostLRoot(LRoot_{pb}, LRoot_{cur}, \sigma_{last})$;\\
				$prog^{\mathbb{E}}.Flush()$; \\
				
				$resolveCensTxs()$;\\	
				$blks_{p}^{\#} \leftarrow 0$ ; \\	
			}
			
		}
		
	}
	\smallskip		
	
	\func{$nextIncProof()$}{
		$LRoot_{tmp} \leftarrow L.add(\textbf{Block}(\perp, [], []))$;\\ 
		$\pi^{inc}_{next} \leftarrow L.IncProof(LRoot_{cur}, ~LRoot_{tmp})$  ;\\
		$L.deleteLastBlock()$;\\
		\textbf{return} $\pi^{inc}_{next}, LRoot_{tmp}$; \Comment{It serves as an incr. proof template for $\mathbb{E}$.}\\
	}
	\smallskip		
	
	\func{$RestoreFailedEnc()$}{
		$hdr_{sync} \leftarrow  L[\#(LRoot_{pb}) ~-1].hdr$; \\
		$blks_{unsync} \leftarrow L[\#(LRoot_{pb}):\text{-}1]$ ; \\
		$L.restore(\#(LRoot_{pb}))$; \Comment{Restore all data to the target version.} \\
		
		$LRoot_{A}, LRoot_{B}, \sigma, PK^{\mathbb{E}}_{pb}, PK^{\mathbb{E}}_{tee} \leftarrow prog^{\mathbb{E}}.ReInit(LRoot_{pb}, blks_{unsync}, hdr_{sync})$; \\
		
		\textbf{assert} $LRoot_{cur} = LRoot_{A}$; \Comment{$\mathbb{E}$ and $\mathbb{O}$ run VM into the same point.}\\
		
		$prog^{\mathbb{S}}.ReplaceEnc(PK^{\mathbb{E}}_{pb}, PK^{\mathbb{E}}_{tee}, LRoot_{A}, LRoot_{B}, \sigma)$; \\
	}
\end{algorithm}

\end{document}